\documentclass[5p, twocolumn]{elsarticle}

\journal{Journal of Systems and Software}

\usepackage{rotating} 
\usepackage{amsmath}
\usepackage{utfsym}
\usepackage{graphicx}
\usepackage{textcomp}
\usepackage{float}
\usepackage[hang,flushmargin]{footmisc}
\usepackage{fourier}
\usepackage{subcaption}
\usepackage{import}
\usepackage{color}

\usepackage{soul}
\usepackage{pifont}
\usepackage{enumitem}
\usepackage{multirow}
\usepackage{tcolorbox}
\usepackage{blindtext}
\usepackage{dirtytalk} 
\usepackage{xspace}
\usepackage{tabularx} 
\usepackage{multirow}
\usepackage{listings} 
\usepackage{todonotes}
\usepackage{booktabs}
\lstdefinestyle{python}{ 
    xleftmargin=6.0ex,
    xrightmargin=2.0ex,
    numbers=left,
    frame=single
}
\usepackage{longtable}
\usepackage{url}

\newcommand{\ReviewerA}[1]{\textcolor{black}{#1}}  
\newcommand{\ReviewerB}[1]{\textcolor{black}{#1}}   

\newenvironment{ReviewerAEnv}{\color{black}}{}
\newenvironment{ReviewerBEnv}{\color{black}}{}

\definecolor{gray10}{gray}{.9}
\definecolor{arsenic}{rgb}{0.23, 0.27, 0.29}
\usepackage{color}

\RequirePackage{fontawesome}

\definecolor{gray50}{gray}{.5}
\definecolor{gray40}{gray}{.6}
\definecolor{gray30}{gray}{.7}
\definecolor{gray20}{gray}{.8}
\definecolor{gray10}{gray}{.9}
\definecolor{gray05}{gray}{.95}

\newlength\Linewidth
\def\findlength{\setlength\Linewidth\linewidth
  \addtolength\Linewidth{-4\fboxrule}
  \addtolength\Linewidth{-3\fboxsep}
}
\newenvironment{examplebox}{\par\begingroup
  \setlength{\fboxsep}{5pt}\findlength
  \setbox0=\vbox\bgroup\noindent
  \hsize=0.95\linewidth
  \begin{minipage}{0.95\linewidth}\normalsize}
  {\end{minipage}\egroup
  \textcolor{gray20}{\fboxsep1.5pt\fbox
    {\fboxsep5pt\colorbox{gray05}{\normalcolor\box0}}}
  \endgroup\par\noindent
  \normalcolor\ignorespacesafterend}

\newenvironment{rqbox}{\par\begingroup
	\setlength{\fboxsep}{5pt}\findlength
	\setbox0=\vbox\bgroup\noindent
	\hsize=0.95\linewidth
	\begin{minipage}{0.95\linewidth}\normalsize}
	{\end{minipage}\egroup
	\textcolor{gray20}{\fboxsep1.1pt\fbox
		{\fboxsep5pt\colorbox{gray05}{\normalcolor\box0}}}
	\endgroup\par\noindent
	\normalcolor\ignorespacesafterend}

\usepackage{color}

\newcounter{Finding}
\stepcounter{Finding}

\begin{document}
\begin{frontmatter}

\title{Evaluating Time-Dependent Methods and Seasonal Effects in Code Technical Debt Prediction}

\author[OULU]{Mikel Robredo}
\ead{mikel.robredomanero@oulu.fi}

\author[LUX]{Nyyti Saarim\"{a}ki}
\ead{nyyti.saarimaki@uni.lu}

\author[OULU]{Matteo Esposito}
\ead{matteo.esposito@oulu.fi}

\author[OULU]{Davide Taibi}
\ead{davide.taibi@oulu.fi}

\author[BIC]{Rafael Pe\~naloza}
\ead{rafael.penaloza@unimib.it}

\author[OULU]{Valentina Lenarduzzi}
\ead{valentina.lenarduzzi@oulu.fi}

\address[OULU]{University of Oulu, Finland}
\address[LUX]{University of Luxembourg, Luxembourg}
\address[BIC]{University of Milano-Bicocca, Italy}

\begin{abstract}
\ReviewerB{
\emph{Background.} Code Technical Debt (Code TD) prediction has gained significant attention in recent software engineering research. However, no standardized approach to Code TD prediction fully captures the factors influencing its evolution.}

\noindent\emph{Objective.} Our study aims to assess the impact of time-dependent models and seasonal effects on Code TD prediction. It evaluates such models against widely used Machine Learning models also considering the influence of seasonality on prediction performance.

\noindent\emph{Methods.} We trained 11 prediction models with 31 Java open-source projects. To assess their performance, we predicted future observations of the SQALE index.  \ReviewerA{To evaluate the practical usability of our TD forecasting model and their impact on practitioners, we surveyed 23 software engineering professionals.}

\noindent\emph{Results.} Our study confirms the benefits of time-dependent techniques, with the ARIMAX model outperforming the others. Seasonal effects improved predictive performance, though the impact remained modest. \ReviewerA{ARIMAX/SARIMAX models demonstrated to provide well-balanced long-term forecasts. The survey highlighted strong industry interest in short- to medium-term TD forecasts.}

\noindent\ReviewerB{
\emph{Conclusions.} Our findings support using techniques that capture time dependence in historical software metric data, particularly for Code TD. Effectively addressing this evidence requires adopting methods that account for temporal patterns.}

\end{abstract}

\begin{keyword}
Technical Debt, Software Quality Mining Software Repositories, Empirical Software Engineering, Time Series Analysis
\end{keyword}

\end{frontmatter}

\section{Introduction}
\label{sec:Introduction}

In Software Engineering (SE), and specially in Mining Software Repository (MSR) studies, researchers often investigate the relationships among different variables collected from the history of software projects. As an example, researchers have investigated the correlations between two variables, such as code smells~\cite{PalombaTSE2018, ArcelliFontana2019, Sharma2020}, their trend over time~\cite{Olbrich2009}, and the impact of different qualities of software~\cite{Sjoberg2013, Palomba2018, lenarduzzi2020some}. 
However, many of these studies have omitted the hidden potential impact on the temporal dependency of the variables analyzed and the possible threats related to statistical techniques not designed for analyzing temporally dependent data~\cite{SaarimakiSANER2022}.   
As an example, the introduction of a technical issue in a commit heavily depends on the code that was present in the repository \emph{before} said commit. However, most studies have not considered this aspect, mainly due to a lack of clear guidelines on the statistical techniques used in this context.  
Saarimäki et al.~\cite{SaarimakiSANER2022} highlighted three main issues in previous studies: 1)~discarding the temporal nature of the commits,  2)~assuming independence of data, and 3)~mixing projects of different sizes, where big projects overwhelm small ones. 
To overcome these issues, they proposed to analyze the dependent data in the SE field, considering the dependency through the time effect~\cite{SaarimakiSANER2022}. 
Within the provided study context, we use the code technical debt (code TD) prediction analysis to evaluate the impact of the time dependence factor. Code TD is an essential metric in software projects as it measures professionals' efforts to clean the code. 
Therefore, code TD denotes the time dependence on past mistakes. Consequently, data analysis techniques that assume this dependence should be considered. 

In this direction, the research community traditionally uses machine learning (ML) models~\cite{tsoukalas2021machine, Tsoukalas2020}. Similarly, we identified two relevant works that applied Time Series Analysis (TSA) to predict code TD~\cite{Mathioudaki2022, Zozas2023} comparing univariate and multivariate TSA models. The results obtained were promising in terms of prediction performance. Hence, our objective is to analyze the robustness of TSA techniques predicting time-dependent variables such as code TD by comparing them with traditional ML methods. Therefore, we are striving to identify existing potential variables that can help explain the behavior of code TD.

We designed and conducted an empirical study to evaluate the code TD prediction performance of different TSA models and a set of time-agnostic ML models commonly used in the literature, which do not consider the data's temporal factor. Furthermore, we assessed the impact of the seasonality factor on the observed data. For that, we adopted the seasonally adjusted version of the TSA models already used in the SE literature and measured their performance to quantify the impact of addressing the seasonality factor. \ReviewerB{As the potential dependent variable adopted for the description of the code TD, we considered the SQALE index metric, computed by SonarQube (SQ) to measure code TD.} 
Specifically, we adopted two different multivariate applications of the \emph{Autoregressive Integrated Moving Average} (ARIMA) model~\cite{box1978analysis} as well as their modified counterparts, which control the seasonality component of the data, and seven ML models to perform the confronted evaluation (four linear and three non-linear). Furthermore, we implemented a backward variable selection approach~\cite{bruce2020practical, james2013introduction} to provide models with the best combination of variables to help improve their efficiency. \ReviewerA{The best resulting multivariate TSA models were afterwards used to explore their accuracy when performing long-term time-series forecasting, a facet of interest in software maintenance activities~\cite{Tsoukalas2020}.}

\begin{ReviewerAEnv}
Finally, to evaluate the practical usability of our TD forecasting model and their impact on practitioners, we surveyed 23 software engineering professionals. The survey was used to determine their willingness to use the model, the usability of the model in real-world TD management, and the predicted window of choice for decision making. The survey asked questions about the professional experience of the respondents, their knowledge of TD and whether they manage TD issues in real life. In addition, the participants provided information on the best prediction horizons, trading off precision for predictions with practicability. The study supports the notion that practitioners desire high-precision predictions in short to medium-term horizons, which is consistent with agile software development cycles and allows for the use of TD predictions within industry pipelines.
\end{ReviewerAEnv}

Our results revealed a robust superiority obtained by the multivariate TSA models and demonstrated a slight improvement in predictive performance when adjusting the models to capture the seasonality effect. Basically, \emph{Seasonally adjusted Autoregressive Integrated Moving Average} with extra regressors (SARIMAX)~\cite{durbin2012time} and its non-adjusted counterpart ARIMAX~\cite{hyndman2018forecasting} were the best prediction models among the methods considered in this study. These both methods implement a multivariate variant of the commonly known ARIMA model. However, the ARIMAX model outperforms the other ML algorithms in predictive performance~\cite{hyndman2018forecasting}. \ReviewerA{Long-term forecast performance showed increased MAPE errors over time but remained balanced across projects. The MAPE error of the SARIMAX model increased from 1.08\% at 6 months to 4.19\% at 36 months, while ARIMAX remained at 3.15\% after 12 biweekly periods and 5.97\% after 72 periods. The variance remained low, with stable trends in descriptive statistics. Finally, our survey reveals good industry confidence in our tool and its effectiveness for short- to medium-term forecasting windows, which aligns closely with industry needs.}

The main contributions of this paper are:

\begin{itemize}
\item An improvement in current knowledge of the prediction of the code TD with an emphasis on investigating the importance of including time-dependent factors. 
\item An additional contribution to the scientific community emphasizing the potential benefits of considering a monitored seasonal code TD measurement for subsequent improvement on the control of code TD in software projects.
\item A large-scale comparison among the time-dependent and ML prediction models already adopted in the literature. 
\item \ReviewerA{A demonstration of the promising capabilities of classic, yet efficient multivariate TSA forecasting models on long-term code TD forecasting, aiming to enable system engineers and project managers to perform long-term effective software maintenance.}
\item \ReviewerA{An industrial survey with 23 experts confirming strong industry interest in short- to medium-term TD forecasts, aligning with agile workflows.}
\end{itemize}

\paragraph{Paper structure} Section~\ref{sec:Background} describes the background on which our paper is based. In Section~\ref{sec:ESDesign}, we present the empirical study design. Section~\ref{sec:Result} presents the results obtained, and Section~\ref{sec:Discussion} shows the derived discussion based on them. Section~\ref{sec:Threats} describes the identified threats to the validity of our study. Section~\ref{sec:RelatedWork} presents related work and Section~\ref{sec:Conclusion} divulges the conclusions of this study and outlines the potential future work.

\section{Background}
\label{sec:Background}

SQ is one of the most commonly adopted static analysis tools both in academia~\cite{LenarduzziICSE2017,LenarduzziSEDA2018} and in industry~\cite{VassalloEMSE19}. 
We aim to help developers create more clean, secure, maintainable, readable, and modular code. SQ evaluates the code by testing it against a predefined set of rules. The rules are considered \emph{coding standards}, and breaking them is deemed undesirable. Therefore, each time a code violates a rule, the SQ presents an issue to the developers. 
Practitioners can use SQ on its official "as a Service" flavor hosted on the sonarcloud.io website or the on-demand version to run on a private server. The tool supports 29 programming languages: Java, Python, C++, and JavaScript.\footnote{https://docs.SQ.org/9.6/analyzing-source-code/languages/overview/} It performs various calculations, including measuring metrics like lines of code and code complexity. It also establishes specific thresholds known as ``quality gates'' for each metric and rule.

In this study, the selected dataset (Section~\ref{sec:Context}) analyzed projects with SQ version 7.5, which includes three rule categories:
\begin{itemize}
    \item \textit{Reliability rules} named as bugs that create an issue that ``represents something wrong in the code'' and that ``will soon be reflected in a bug in the code.'' 
    \item \textit{Maintainability rules}  named as code smells that decrease code readability and modifiability. It is important to note that the term ``code smells'' adopted in SQ contains only a subset of the well-known code smells defined by Fowler et al.~\cite{Fowler1999}.  
    \item \textit{Security rules} named as Vulnerability, creates a problem that impacts the application's security. 
\end{itemize}

\ReviewerA{To quantify and therefore measure the existing technical debt in software projects, SQ leverages a series of metrics, mostly quantitative translations of code analysis metrics. In the version considered for this study, SQ computed three types of TD, \textit{reliability remediation effort} and \textit{security remediation effort}, which consisted on the time to fix all the open issues classified as bugs and vulnerabilities, and the \textit{technical debt ratio}, which SQ named as \textit{SQALE index} resembling the naming of SQALE methodology~\cite{SQUALE} used in older versions, and consisting on the time to fix all open issues classified as \textit{code smell}.}

\section{The Empirical Study Design}
\label{sec:ESDesign}
We now describe our study, reporting the goal and research questions, the context, the collection and analysis of data. We designed our experimental study based on the guidelines defined by Wohlin et al. \cite{Wohlin2000}. 

\subsection{Goal and Research Questions}
\label{sec:Goal}
We formalized the goal of this study according to the GQM approach~\cite{Basili1994} as follows:

\vspace{2mm}
\noindent \textit{Analyze} time-dependence and seasonality effect factors, \\
\textit{for the purpose of}  comparing code TD prediction,\\
\textit{with respect to} performance,\\
\textit{from the point of view of}  developers,\\
\textit{in the context of} open source software.

\vspace{2mm}
Based on this goal, our formulated Research Questions (\textbf{RQs}) are presented next.
\begin{center}	
	\begin{rqbox}
		\textbf{RQ$_1$.} \textit{Which multivariate time-dependent method predicts code TD with the highest performance?}
  \end{rqbox}	 
\end{center}
%



\textbf{RQ$_1$} compares two different multivariate TSA approaches already implemented in the SE literature to determine which approach can predict code TD better. Evaluating the prediction performance of the latest models that control the time dependence factor within the data enables us to determine which of these models should be considered by practitioners when deliberating which prediction techniques to use when predicting code TD. To address RQ$_1$, we collected as code TD the SQALE index metric computed by SQ (as explained in Section~\ref{sec:Background}), and Code Smell rule violations generated by SQ as potential explanations of the SQALE index's behavior.

The first approach, known as ARIMA + LM, combines multiple ARIMA models and linear regression models (LM)~\cite{Zozas2023}. This approach utilizes a univariate ARIMA model for the prediction of each independent variable. Using predicted values within a regression model, it predicts future values of the dependent variable. The second approach, known as ARIMAX, consists of building a multivariate ARIMA model to predict the values of the dependent variable~\cite{Mathioudaki2022}. In the ARIMAX model, the history of the dependent variable and the past values of the considered independent variables are used as features to train the model and provide future predictions of the dependent variable.

Our prior assumption is that the ARIMAX model performs better than the ARIMA + LM models in explaining the future evolution of the dependent variable. We assume that combining the independent variables with the past values of the dependent variable has more information than separately predicting the values of the independent variables to predict the dependent variable through a regression model. 

The purpose of this study is to establish the applicability of the most efficient time-dependent techniques for code TD prediction and evaluate how the adopted techniques perform against other prediction techniques commonly used in the SE literature. These techniques are mostly ML-based algorithms that do not consider the temporal factor when providing future predictions. Therefore, we formulate our \textbf{RQ$_2$}.


\begin{center}	
	\begin{rqbox}
		\textbf{RQ$_2$.} \textit{How accurate is the prediction performance of a multivariate time-dependent approach for code TD prediction compared to that of an ML algorithm?}
  \end{rqbox}	 
\end{center}

To assess the applicability of a time-dependent approach for code TD prediction (\textbf{RQ$_2$}), in our case multivariate TSA,  we need to compare its performance with other approaches that do not consider the data's time dependency. In this study, the comparison is made with several ML models.

Similarly, as for \textbf{RQ$_1$}, our prior assumption is that the prediction performance of multivariate TSA methods for code TD prediction overcomes that of ML models. Moreover, TSA methods are based on temporally ordered serialized data, helping the models capture potential temporal factors, for example, the seasonality of the observation~\cite{chan2008time}. Models can be adjusted to these factors to observe their impact on the predicted results. Therefore, only comparing already used TSA methods is not sufficient, and we formulate a third research question.

\begin{center}	
	\begin{rqbox}
		\textbf{RQ$_3$.} \textit{Does addressing seasonality improve the prediction performance of multivariate time-dependent approaches for code TD prediction?}
    \end{rqbox}	 
\end{center}

As explained, \textbf{RQ$_3$} expands \textbf{RQ$_1$}. For that, given the modification of the models to handle seasonality patterns,
\textbf{RQ$_3$ }compares the new models with the former in terms of the prediction of the code TD and therefore evaluates the impact of seasonality in the prediction performance.

Our prior assumption is that SARIMAX, SARIMA + LM approaches and seasonally adjusted models of the previously defined TSA model counterparts perform better than their former model format, which did not handle seasonality patterns. In addition, we take advantage of the comparison proposed in RQ$_{1}$. Thus we assume that using the multivariate SARIMAX model is more informative than the SARIMA+LM model, given the extension of both TSA models to capture the impact of the seasonality effects.

Similarly, evaluating the seasonality adjustment in prediction performance compared to the selected ML algorithms becomes paramount to contributing to the SE community through this study. Therefore, we ask a fourth question.

\begin{center}	
	\begin{rqbox}
		\textbf{RQ$_4$.} \textit{How accurate is the prediction performance of a seasonally adjusted multivariate time-dependent approach for code TD prediction compared to that of an ML algorithm?}
    \end{rqbox}	 
\end{center}

Our prior assumption is that the additional modification implemented in the multivariate TSA models regarding seasonality control enables their code TD prediction performance to be better than that of ML models. 

For all RQs described so far, we evaluated the models using three key \textbf{performance metrics}: Mean Absolute Percentage Error (MAPE), Mean Absolute Error (MAE), and Root Mean Square Error (RMSE) (see Section~\ref{sec:performancemet} for a detailed description of these metrics).

\begin{ReviewerAEnv}
\begin{center}	
	\begin{rqbox}
		\textbf{RQ$_5$.} \textit{\ReviewerA{How accurate are multivariate time-dependent methods on long-term forecasting?}}
    \end{rqbox}	 
\end{center}

\ReviewerA{A challenge that emerges from the concept of code TD prediction is the need to make multi-step forecasts, that is, forecasts for more than one time-step into the future. Providing reasonable long-term forecasts would allow system engineers and project managers to perform long-term effective software maintenance. While single-step predictions can be useful for short-term planning, multi-step forecasting offers more practical insights for software maintenance and allows project managers to make strategic decisions for the future. Therefore, through the defined \textbf{RQ$_5$} we will investigate the performance of the resulting best TSA models when performing multi-step forecasting to enable practitioners observe the long-term impact of their current project maintenance level within the context of code TD.}

\begin{center}	
	\begin{rqbox}
		\textbf{RQ$_6$.} \textit{\ReviewerA{How valuable do software practitioners perceive a Technical Debt forecasting to be?}}
\begin{itemize}
    \item \textbf{RQ$_{6.1}$.} \textit{\ReviewerA{What forecasting window combinations do practitioners consider acceptable and useful for Technical Debt prediction?}}
    \item \textbf{RQ$_{6.2}$.} \textit{\ReviewerA{What is the preferred scope of Technical Debt prediction among practitioners?}}
\end{itemize}        
    \end{rqbox}	 
\end{center}

While many techniques exist to measure and manage code TD retrospectively, forecasting its evolution could allow practitioners to take proactive measures.
Understanding how practitioners perceive code TD forecasting is crucial to determining whether such predictive models can influence decision-making processes in software projects or are just academic interests (\textbf{RQ$_6$}). If practitioners do not see substantial value, efforts in this direction might need re-evaluation or refinement to align with industry needs.

To further explore this aspect, we investigate the specific forecasting configurations that practitioners deem useful.
%
To effectively forecast the evolution of code TD (\textbf{RQ$_{6.1}$}), it is crucial to determine the appropriate time horizon for predictions. Practitioners may prefer short-term forecasts over weeks or months to aid in immediate refactoring efforts. In contrast, long-term forecasts, spanning years, could be more useful for strategic planning.
Knowing which time frames practitioners consider valid and useful ensures that code TD prediction models meet actual industry needs, thus facilitating their practical application.
Moreover, selecting suitable forecast windows is just one part of the equation. Defining the preferred scope of code TD predictions to customize these forecasts properly is equally important.


Code TD can appear at different levels, from specific code components to software projects. Some practitioners focus on predicting TD at the module or class level to aid in detailed refactoring efforts (\textbf{RQ$_{6.2}$}). In contrast, others prefer predictions at the system-wide level to help with architectural decisions. Understanding the preferred scope ensures that code TD forecasting tools provide valuable insights at the appropriate level of abstraction, making them more useful in practitioners' daily workflows. Our survey offers a comprehensive view on creating Technical Debt forecasting methods that are practical and actionable for software professionals.
\end{ReviewerAEnv}


\ReviewerA{For \textbf{RQ$_5$} and \textbf{RQ$_6$}, we evaluated the models using only MAPE since it is one of the most intuitive metrics for practitioners to discuss~\cite{carka_effort-aware_2022,esposito2024large}.}

\subsection{Context}
\label{sec:Context}

In the context of our study, we consider the Technical Debt Dataset~\cite{LenarduzziPromise2019} version~2.0 adopted by Saarimäki et al.~\cite{SaarimakiSANER2022}. The data set contains 31 Java projects from the Apache Software Foundation (ASF) repository.\footnote{http://apache.org} The projects in the data set were selected based on ``criterion sampling''~\cite{Patton2002}, requiring all of the following criteria: developed in Java, older than 3 years, more than 500 commits and 100 classes, and the usage of an issue tracking system with at least 100 issues reported. 
The projects were selected, maximizing their diversity and representation by considering a comparable number of projects on the age, size, and domain of the project. The projects can be considered mature due to the strict review and inclusion process required by the ASF. In addition, the included projects regularly review their code and follow a strict quality process.\footnote{https://incubator.apache.org/policy/process.html} Full details of the data can be found in the online repository of the data set.\footnote{https://github.com/clowee/The-Technical-Debt-Dataset}
We adopted this data set since all its projects were already analyzed with SQ and it contains the data related to the SQALE index.

\subsection{Variables} 
\label{sec:variables} 
In this section, we describe the dependent and independent variables used in the models adopted for the prediction of code TD.

\begin{itemize} [leftmargin=*]
    \item \textbf{Dependent variable}: We consider the metric SQALE index computed by SQ. \textit{The SQALE index} denotes the efforts to fix all the Code Smell rules that are violated in the code and is measured in minutes. SQ denotes TD using the total value of the SQALE index. The metric is continuous and, therefore, is suitable for the TSA models adopted in this study. In addition, it provides a clear definition of code TD to answer the objective of our study.
    
    \item \textbf{Independent variables:}  We consider the Code Smell rule violations detected by SQ. \textit{Code smell rule violations} describe the number of violations for each SQ rule of code smell type. The TD dataset has issues from 205 different rules of type code smell, and each rule is included as a separate independent variable in the study. The nature of the collected issues is discrete as it is based on the count of issues detected in each SQ analysis for each code smell type.
\end{itemize}

\subsection{Data Collection}
\label{sec:DataCollection}

\begin{ReviewerBEnv}
In this study, we were interested in studying the impact of temporal factors on the selected dependent variable, the SQALE index. As described in Section 2, code smell rule violations are directly related to the SQALE index, which is the remediation cost. Hence, we decided to collect the 205 code smell rule violations existing in the adopted data set as the initial set of independent variables for our study.
\end{ReviewerBEnv}

\subsubsection{Data preprocessing: Feature selection}

The data collected contains 205 potential independent variables. Such a large number could have a negative impact on the models due to the so-called ``curse of dimensionality:'' including a large number of independent variables may drastically reduce the predictive power of a model. 
To identify a set of the most explicative independent variables for our dependent variable, we undertook a feature selection process based on well-known feature selection techniques. The aim was to categorize the potential independent variables according to their importance. These techniques are briefly described next.

\begin{figure*}[htb]
    \centering
    \includegraphics[width=\linewidth]{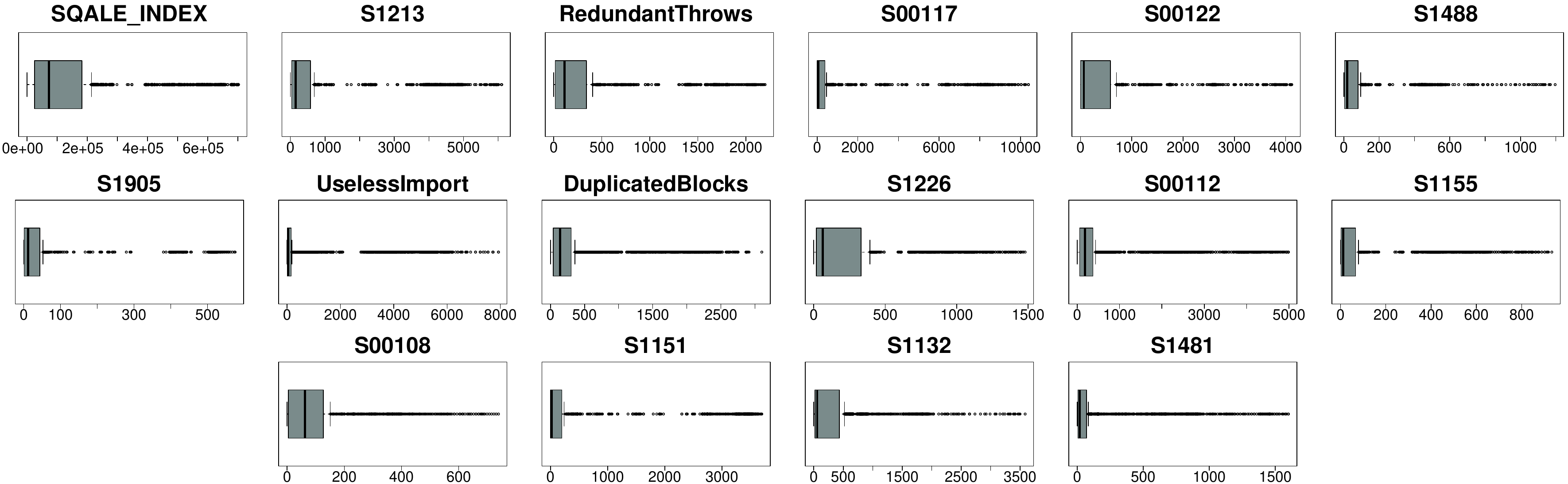}
    \caption{Boxplots for the considered model variables.}
    \label{fig:variables_boxplot}
\end{figure*}

\begin{itemize} [leftmargin=*]
    \item \textbf{Variance Thresholding}: Method that utilizes the variance or spread of a variable in a data set to select model-independent variables. In summary, this method calculates the variance of all potential independent variables in the data sets. Removes those with low variance as they are less informative and do not have much predictive power~\cite{fida2021variance}.

    \item \textbf{Zero percentage method}: This method eliminates predominantly empty independent variables, i.e. the percentage of emptiness is higher than a predefined threshold and is therefore poorly informative. It is used with count data~\cite{allison2009missing}.

    \item \textbf{Feature importance}: This method determines the importance of each variable in a data set using a Random Forest~(RF) algorithm. RF is an ensemble learning technique consisting of multiple decision trees, in this case, used for prediction. The measure of importance is based on the reduction in node impurities due to the splitting of the variable. Averaged across all trees, the measure of the impurity of the node of all independent variables is calculated by the residual sum of squares. Thus, this method excludes variables with low feature importance~\cite{speiser2019comparison}.

    \item \textbf{Correlation analysis}: The method evaluates the degree of linear association between the independent and dependent variables. Given the high skewness illustrated in Figure~\ref{fig:variables_boxplot}, and therefore lack of normality in our variables~\cite{agresti2015foundations}, we measure the correlational association through the non-parametric Spearman's \textit{$\rho$} rank correlation coefficient~\cite{spearman1961proof} for each independent variable. Thus, variables with a poor informative contribution for predictive modeling are discarded~\cite{yu2003feature}. 
    \end{itemize}

    
One could expect that each of the selected feature selection techniques would provide different results on the initial collected issue-type variables. Therefore, we ranked the independent variables in terms of their resulting importance score for each of the importance techniques of the characteristics, resulting in four different rankings. We used the \textit{Interquartile Range} (IQR) method to select the upper quartile in the ranking distribution based on the importance scores. In a nutshell, given a set of observations, the IQR method builds the underlying distribution from the observations. Select the subset of the different quartiles in the distribution based on the objective of the analysis~\cite{kim2021choosing}. In our scenario, we were interested in selecting the most informative independent variables of our set of variables; therefore, we selected the quartile with the variables denoting the highest feature importance score. Additionally, for the sake of consistency in the final set of independent variables, we selected the subset of variables that showed importance in each of the four fetched quartiles.
Consequently, the feature selection process resulted in a subset of 15 independent variables that were specifically selected to predict the SQALE index (dependent variable). Table~\ref{tab:modelVariables} and Figure~\ref{fig:variables_boxplot} provide a graphical and quantitative representation of the independent variables. 

\subsubsection{Data preprocessing: Transforming raw data into time series data}
\label{sec:preprocessing_timeseries}

The time-dependent approaches used in this paper require the use of periodically measured time-serialized data. Given that commits are not periodic, we relied on two criteria to generate the time series. First, we focused on constructing a sufficiently long time series for each project (study subjects) to provide the models with sufficient data to perform the prediction. Second, we investigated different periodicity levels for the observations in the time series. In this way, we could observe the temporal nature of the data and find the most suitable time frame between observations.  

Based on the described criteria, we generated time series data at biweekly and monthly periodicity levels. The number of data points for each project on each periodicity level is presented in Table~\ref{tab:descriptiveStats}. TSA models require at least 24 observations to detect existing time-dependent factors such as the temporal trend, and consequently to be trained~\cite{chan2008time}. Therefore, increasing the size of the time window would reduce the number of observations for the collected projects and would require excluding the projects from the study.

Version control data are stochastically ordered by nature,\footnote{https://www.git-scm.com/docs/git-commit} which means that some time periods (of weeks or months in our study) might not have commits despite being otherwise an active project. Consequently, serializing such data can result in missing data for some time points.  Linear interpolation was used to fill in time points that lacked observations in the generated time series. This method presupposes a linear correlation between neighboring data points and calculates the values of the missing data points using the known values surrounding them~\cite{blu2004linear}.

Consequently, the analysis was performed using two different samples of the data: \textit{monthly} time series data and \textit{biweekly} time series data. As a potential threat to the validity of this study, we address the interpolation of artificially generated observations in Section~\ref{sec:Threats}. Similarly, we provide suggestions for practitioners on the remediation approach to avoid missing data in Section~\ref{sec:descriptivestats} to help practitioners generate higher quality data.

\begin{table*}[htb]
    \caption{Descriptive statistics of the SQALE index and the resulting independent variables from the data preprocessing stage.}
    \label{tab:modelVariables}
    \centering
    \footnotesize
    \begin{tabular}{@{}l|rr|rrrrrr@{}}
        \hline
        \textbf{Variable} & \textbf{Mean} & \shortstack{\textbf{Standard} \\ \textbf{deviation}} & \textbf{Min} & \shortstack{\textbf{Lower} \\ \textbf{quartile}} & \shortstack{\textbf{Median} \\ \textbf{value}} & \shortstack{\textbf{Upper} \\ \textbf{quartile}} &\textbf{Max} & \textbf{Skewness}\\
        \hline
        SQALE index & 137,726 & 169,927 & 0 & 24,378 & 72,981 & 182,132 & 701,914 & 1.87\\
        S1213 & 740 & 1,350 & 0 & 38 & 152 & 579 & 6,106 & 2.22\\
        RedundantThrows DeclarationCheck & 345 & 509 & 0 & 15 & 113 & 340 & 2,200 & 1.91\\
        S00117 & 1,082 & 2,380 & 0 & 7 & 55 & 383 & 10,395 & 2.3\\
        S00122 & 402 & 658 & 0 & 1 & 67 & 585 & 4,119 & 2.55\\
        S1488 & 88 & 163 & 0 & 4 & 20 & 81 & 1,196 & 2.85\\
        S1905 & 74 & 148 & 0 & 1 & 12 & 44 & 575 & 2.26\\
        UselessImportCheck & 630 & 1,402 & 0 & 7 & 42 & 151 & 7,928 & 2.58\\
        DuplicatedBlocks & 333 & 510 & 0 & 40 & 142 & 305 & 3,099 & 2.52\\
        S1226 & 212 & 293 & 0 & 18 & 62 & 330 & 1,479 & 1.91\\
        S00112 & 473 & 848 & 0 & 58 & 183 & 374 & 4,987 & 3.01\\
        S1155 & 86 & 169 & 0 & 2 & 12 & 67 & 933 & 2.49\\
        S00108 & 97 & 128 & 0 & 4 & 62 & 127 & 739 & 1.95\\
        S1151 & 472 & 1,007 & 0 & 2 & 26 & 200 & 3,699 & 2.25\\
        S1132 & 289 & 481 & 0 & 20 & 57 & 436 & 3,587 & 2.94\\
        S1481 & 132 & 274 & 0 & 5 & 18 & 71 & 1,600 & 2.75\\
        \hline
    \end{tabular}
    
\end{table*}

\begin{ReviewerAEnv}
\begin{table*}[htb]
    \centering
    \centering
    \caption{\ReviewerA{Survey: Questions and research questions (RQ$_6$)\\{\footnotesize Legend: \textbf{C} - Closed, \textbf{C*} - Closed with ``Other'' option,  \textbf{L} - Linkert, \textbf{O} - Open Ended}}}
    \small
    
    \label{tab:technical_debt_survey}
    \resizebox{\linewidth}{!}{%
    \begin{tabular}{@{}l| p{0.2cm} p{16.5cm}| l@{}}
    \hline
    \textbf{RQ }&  & \textbf{Question (Q)} & \textbf{Type} \\
    \hline
    \multirow{11}{*}{\rotatebox{90}{Profiling}} & Q$_1$ & What is your job title or role? & O \\ 
     & Q$_2$ & What sector does your organization belong to? & O \\ 
     & Q$_3$ & What is the size of your organization? & C \\ 
     & Q$_4$ & Where is your organization located? & C \\ 
     & Q$_5$ & How many years of experience do you have? & C \\ 
     & Q$_6$ & What types of projects do you usually work on? & C* \\ 
     & Q$_7$ & Do you use any agile development practices? & C \\ 
     & Q$_8$ & Are you familiar with the concept of technical debt? & C \\ 
    &  Q$_9$ & If yes, what agile practice do you employ? & O \\ 
    &  Q$_{10}$ & How often do you address Technical Debt-related issues? & O \\ 
     & Q$_{11}$ & What are the types of Technical Debt issues you address most frequently? & O \\ \hline
    \multirow{6}{*}{\rotatebox{90}{RQ$_{6.1}$}} &Q$_{12}$ & How much would you value a tool for Technical Debt forecasting? & L \\ 
   &   Q$_{13}$ & Please justify your choice in the previous question.  & O \\ 
    &  Q$_{14}$ & Which forecasting window combinations are acceptable and useful for your workflow? & C \\ 
    &  Q$_{15}$ & Please justify your choice in the previous question. & O \\ 
    &  Q$_{16}$ & Among the following options, what forecasting window combinations may you find acceptable and useful for your workflow? & C \\
    & Q$_{17}$ & Please justify why you chose the previous options. & O \\ \hline
    \multirow{2}{*}{\rotatebox{90}{RQ$_{6.2}$}} &  Q$_{18}$ & If given the choice, would you rather choose to predict Technical Debt: Weekly, By-Weekly, Monthly, Other? & C* \\ 
    &  Q$_{19}$ & Please justify your choice in the previous question.  & O \\ \hline
\end{tabular}%
}
\end{table*}

\subsection{\ReviewerA{Collecting the practitioners' forecasting perceived valuable}}

TD is a persistent challenge in software development, affecting maintainability, quality, and long-term sustainability. Although retrospective code TD analysis is widely studied, proactive forecasting remains an emerging area with limited understanding of its perceived value in industry. To bridge this gap, we designed this survey to capture the perspectives of software professionals on the usefulness, preferred forecasting windows, and optimal scope of Code TD predictions.
Our target audience consists of software engineers, architects, and technical leaders involved in technical debt management. Our objective is to evaluate whether our forecasting model is perceived as beneficial to the industry. Specifically, we want to determine the most advantageous forecasting time windows that add value to the software development lifecycle (SDLC) within an industrial setting.

\subsubsection{Population Description}
\begin{table}[htb]
    \centering
\footnotesize
    \caption{\ReviewerA{Interviewees' Professional Experience - Part 1}}
    \label{tab:demographic_part_one}
    \begin{tabular}{l l p{4cm} r}
        \hline
         & \textbf{Question} & \textbf{Response} & \textbf{\%} \\ \hline
        \multirow{3}{*}{Q$_1$} & \multirow{3}{*}{Job title} & Software Architect & 9  \\ \cline{3-4}
                            &                            & Software Developer & 30  \\ \cline{3-4}
                            &                            & Software Engineer & 4  \\ \hline
        \multirow{5}{*}{Q$_2$} & \multirow{5}{*}{Organization sector} & Cloud Architect & 4  \\ \cline{3-4}
                            &                                    & ICT & 35  \\ \cline{3-4}
                            &                                    & IT Consulting & 30  \\ \cline{3-4}
                            &                                    & Software Engineering & 26  \\ \cline{3-4}
                            &                                    & Software Development & 4  \\ \hline
        \multirow{4}{*}{Q$_3$} & \multirow{4}{*}{Organization size} & Large Enterprise & 65  \\ \cline{3-4}
                            &                                    & SME & 17  \\ \cline{3-4}
                            &                                    & Small & 13  \\ \cline{3-4}
                            &                                    & Start-up & 4  \\ \hline
        \multirow{3}{*}{Q$_4$} & \multirow{3}{*}{Organization location} & Europe & 87  \\ \cline{3-4}
                            &                                        & North America & 4  \\ \cline{3-4}
                            &                                        & South America & 9  \\ \hline
        \multirow{5}{*}{Q$_5$} & \multirow{5}{*}{Years of experience} & < 1 & 9  \\ \cline{3-4}
                            &                                      & 1-3 & 26  \\ \cline{3-4}
                            &                                      & 4-7 & 30  \\ \cline{3-4}
                            &                                      & 8-10 & 13  \\ \cline{3-4}
                            &                                      & > 10 & 22  \\ \hline
        \multirow{7}{*}{Q$_6$} & \multirow{7}{*}{Type of Project} & AI/ML & 29  \\\cline{3-4}
                            &                                              & Web Development & 29  \\\cline{3-4}
                            &                                              & Cloud \& DevOps & 12  \\\cline{3-4}
                            &                                              & IoT & 8  \\\cline{3-4}
                            &                                              & Embedded Systems & 8  \\\cline{3-4}
                            &                                              & Mobile Development & 4  \\\cline{3-4}
                            &                                              & Software Architecture \& Quality & 8  \\\hline
    \end{tabular}%
\end{table}

\begin{table}[htb]
    \centering
\footnotesize
    \caption{\ReviewerA{Interviewees' Professional Experience - Part 2}}
    \label{tab:demographic_part_two}
    \resizebox{\linewidth}{!}{%
    \begin{tabular}{l l p{3.5cm} m{1cm}}
        \hline
         & \textbf{Question} & \textbf{Response} & \textbf{\%} \\ \hline
        \multirow{2}{*}{Q$_7$} & \multirow{2}{*}{Use of agile practices} & No & 26  \\ \cline{3-4}
                            &                                       & Yes & 74  \\ \hline
        \multirow{3}{*}{Q$_8$} & \multirow{3}{*}{Familiarity with TD} & No & 13  \\ \cline{3-4}
                            &                                                 & Somewhat & 17  \\ \cline{3-4}
                            &                                                 & Yes & 70  \\ \hline
        \multirow{8}{*}{Q$_9$} & \multirow{8}{*}{Agile Practice Employed} & Scrum (SCRUM, Lean + Scrum) & 56.25  \\ \cline{3-4}
                            &                                                 & CI/CD, Version Control, Test Automation & 18.75  \\ \cline{3-4}
                            &                                                 & Kanban & 6.25  \\ \cline{3-4}
                            &                                                 & MLOps & 6.25  \\ \cline{3-4}
                            &                                                 & Feature-Driven Development (FDD) & 6.25  \\ \hline
        \multirow{8}{*}{Q$_{10}$} & \multirow{8}{*}{Frequency addressing TD} & Weekly & 4  \\ \cline{3-4}
                            &                                                 & Per Sprint & 13  \\ \cline{3-4}
                            &                                                 & Monthly & 13  \\ \cline{3-4}
                            &                                                 & Quarterly & 4  \\ \cline{3-4}
                            &                                                 & Project Start & 4  \\ \cline{3-4}
                            &                                                 & As Needed & 9  \\ \cline{3-4}
                            &                                                 & Occasionally & 22  \\ \cline{3-4}
                            &                                                 & Rarely & 13  \\ \hline
        \multirow{10}{*}{Q$_{11}$} & \multirow{10}{*}{TD type addressed} & Refactoring (general \& systematic) & 26.09  \\ \cline{3-4}
                            &                                                 & Testing Debt (test smells, lack of unit tests) & 17.39  \\ \cline{3-4}
                            &                                                 & Code smells & 13.04  \\ \cline{3-4}
                            &                                                 & Bug fixing (small, unforeseen, product bugs) & 13.04  \\ \cline{3-4}
                            &                                                 & Architectural Debt (smells, poor planning, scalability issues) & 13.04  \\ \hline
    \end{tabular}%
    }
\end{table}

Table~\ref{tab:demographic_part_one} and Table~\ref{tab:demographic_part_two} present the professional experience of our the 23 interviewed experts. The surveyed population is predominantly composed of software professionals, particularly Software Developers, and is heavily concentrated in Europe. The respondents largely work in established firms within the ICT, IT Consulting, and Software Engineering sectors. They possess a range of professional experience, with a significant number employing Agile methodologies in their development practice and familiarity with code TD. Hence, our population appears to be a fair sample of the intended audience.
 We collected the answers anonymously and provided all participants with information regarding the GDPR act for data collection and protection. We also thoroughly followed the ACM publications policy on research involving human participants and subjects \cite{ACM_Human_Participants_2021}.

\subsubsection{Questionnaire}
Table \ref{tab:technical_debt_survey} presents the questions used to answer RQ$_6$. Due to space constraints, we provide the complete questionnaire in the replication package. It should be noted that, before submitting the questionnaire, we have performed a pilot questionnaire to derive the answers to the predefined closed-ended questions (C) predefined answers. Moreover, for most closed-ended questions, we provided an open-ended version~(O) to account for possible missing categories or to allow for an in-depth explanation of the selection. We interviewed experts involved in our previous research and, leveraging domain knowledge, reached a consensus on the predefined answer to closed-ended questions. We did not ask the pilot's experts to participate in the final questionnaire to avoid biases.

According to our guidelines \cite{wholin2012experimentation, DBLP:journals/ese/RunesonH09,Basili1994,rios2020practitioners}, we asked participants to answer demographic questions to obtain information about the population under examination. 
Thus, Q$_1$ to Q$_{11}$ refer to the personal background and professional activities of the interviewee. We asked participants to respond with predefined choices to facilitate data analysis using domain knowledge \cite{esposito2024leveraging,esposito2024large}.

For RQ$_{6.1}$, we asked them for their perceived benefit in predicting the code TD. More specifically, we provide a Likert scale question (L), Q$_{12}$ surveying how much would they value a tool for code TD prediction. A Likert scale is a psychometric scale commonly used in questionnaires to gauge respondents’ attitudes, opinions, or perceptions on a particular topic \cite{likert1932technique}. It typically consists of a series of statements in which respondents indicate their level of agreement or disagreement on a symmetric agree-disagree scale, usually ranging from ``strongly agree'' to ``strongly disagree.'' This method allows the quantitative measurement of people's attitudes or feelings toward a subject. The values for each Likert scale question are available in the replication package. Furthermore, we asked the interviewee to justify the rationale for their Likert choice in Q$_{13}$. In addition, we also provide two closed-ended questions, Q$_{14}$ and Q$_{16}$ to choose between pairs of ``\textit{<time windows, MAPE>}.'' To allow the practitioner to grasp our ability to predict Code TD, we decided, based on RQ$_{5}$ findings, to select a maximum time frame of 36 months, that is, 3 years. We divided such time frames into biweekly prediction time windows Q$_{13}$ and monthly time window Q$_{15}$ by pairing the time window with its computed MAPE (e.g., 2 weeks - 1.44 \%). Since we were interested in understanding the rationale for the choice, we also asked the practitioners to further disclose their motivation for the chosen intervals in the two connected open-ended questions Q$_{15}$ and Q$_{17}$.

Finally, for RQ$_{6.2}$, we ask whether the interviewee prefers to predict the code TD weekly or monthly Q$_{18}$ based on previous comparisons of time windows and MAPE, and to justify their choice in Q$_{19}$.

\end{ReviewerAEnv}

\subsection{Data Analysis}
\label{sec:DataAnalysis}

\begin{figure*}[htb]
    \centering
    \includegraphics[width=\linewidth]{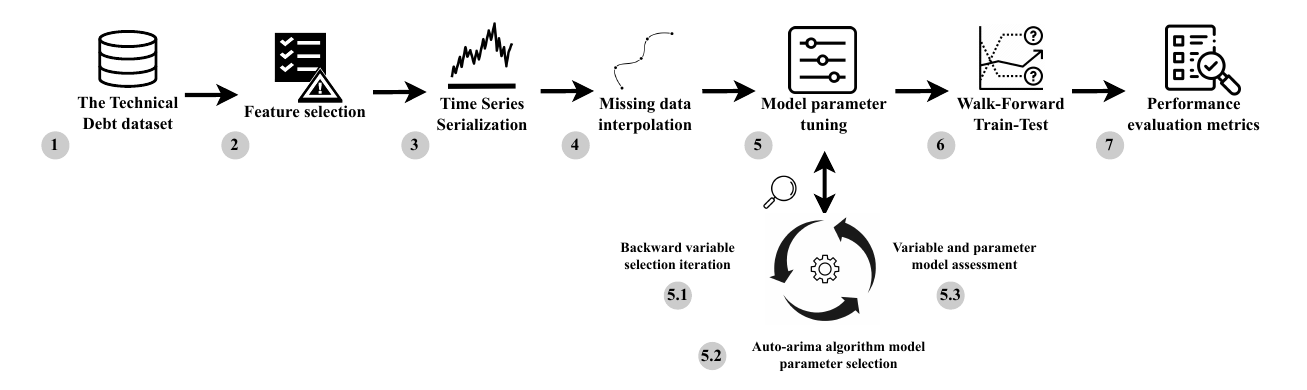}
    \caption{Data analysis process workflow diagram.}
    \label{fig:study_diagram}
\end{figure*}

In this section, we focus on the designed data analysis based on the definition of the selected models, and we explain the comparison of the seasonally adjusted models against their root model counterparts, their model parameter tuning structure,  and the ML models adopted. The selected time-dependent models are based on \textit{multivariate} TSA modeling, a field of statistical modeling commonly used in diverse fields such as econometrics and weather forecasting, for instance~\cite{box2015time}. The multivariate nature of the models allows us to consider information about additional independent variables that can assist the model in the prediction process. \ReviewerA{Figure~\ref{fig:study_diagram} shows the visual representation of the data analysis process performed in this study.}

\subsubsection{Setting the stage for seasonally unadjusted TSA model prediction (RQ$_1$ - RQ$_5$)}
\label{sec:ARIMA}

We examine two distinct methods to predict code TD utilizing TSA models, ARIMAX and ARIMA+LM. As version control data is inherently \textit{non-stationary}, i.e. denote a trend across time, the two methods originate from the \textit{Box-Jenkins} model. better known as \textit{Auto-Regressive Integrated Moving Average}~(ARIMA) model. It is worth noticing that a time series of data is \textit{stationary} when its predictions do not depend on factors such as trend or seasonality existing in the real data~\cite{brockwell1991time}. The ARIMA model is a modified version of the \textit{Autoregressive Moving Average}~(ARMA) model used for modeling non-stationary time series~\cite{dickey1979distribution}. Like the ARMA model, ARIMA determines the best autoregressive parameter for the variable to be predicted (AR step) and also identifies the optimal moving average parameter (MA)~\cite{box2015time}. In addition, ARIMA executes the integration step~(I), which determines the level of differencing required for the time series of the variable in question to achieve stationarity. The methods considered are the following.

\begin{itemize}[leftmargin=*]

    \item \textbf{ARIMAX}: The approach used in~\cite{Mathioudaki2022} was an ARIMA model with a set of independent variables included in the model, commonly referred to as ARIMAX. This model offers a further multivariate approach to the ARIMA model by introducing additional independent variables into the model and, therefore, helping to explain the evolution of the SQALE index. This enables future SQALE index values to be explained both by its past values and by those of the independent variables. 

    \item \textbf{ARIMA + LM}: An ensemble technique previously adopted in~\cite{Zozas2023} that comprises constructing a univariate ARIMA model for each independent variable in the model. Each ARIMA model considers the past values of their respective model variable as model parameters and therefore independently predicts the future values of each model variable. The additional step in this methodology relies on the performance of \textit{ linear regression} (LM) considering each of the new prediction lag values to predict the new values for the SQALE index variable.

\end{itemize}

\subsubsection{Setting the stage for Seasonally Adjusted TSA model prediction (RQ$_{3, 4}$)}
\label{sec:sarima}

In contrast with the previously presented ARIMAX and ARIMA+LM models, the seasonally adjusted models capture the impact of the seasonality pattern in the data while maintaining the model nature of the previously defined TSA model counterparts. Therefore, the analyzed seasonally adjusted methods are as follows.

\begin{itemize}[leftmargin=*]
    \item \textbf{SARIMAX}: The \textit{Seasonally adjusted Auto-Regressive Integrated Moving Average} model stands as the seasonally adjusted extension of the ARIMAX model. Addresses the seasonality of the data within model training through model parameter tuning~\cite{box1978analysis}. More information on the adjustment of the model parameters in the TSA model is provided in Section~\ref{sec:parameterTuning}. Similarly, the SARIMAX model provides a multivariate approach compared to its univariate counterpart, SARIMA, by including independent variables in the model. This enables future SQALE index values to be explained both by its past values and by those of the included independent variables. 

    \item \textbf{SARIMA + LM}: The approach is characterized by the combination of the univariate \textit{ seasonally adjusted integrated moving average} (SARIMA) model combined with a \textit{ linear regression} model (LM). Based on the seasonality adjustment performed in the approach used in~\cite{Zozas2023}, a SARIMA model is built for each of the independent variables considered in the study, and therefore each model independently predicts future values of a single variable accordingly. The additional step in this methodology relies on the implementation of \textit{ linear regression} (LM) considering each new prediction lag value of the independent variables to predict the new values for the SQALE index variable.

\end{itemize}

\subsubsection{\textit{Setting the model parameter order in TSA models (RQ$_1$ - RQ$_5$)}}
\label{sec:parameterTuning}

Similarly to other families of predictive analysis models, for instance, Linear Models or Generalized Linear Models~\cite{agresti2015foundations}, TSA models are founded on a set of model parameters that are estimated after the model is trained. As mentioned earlier, in this study we adopt as prediction models derivations from the \textit{Box-Jenkins} ARIMA model~\cite{box1978analysis}. Models founded on the ARIMA model follow the same model parameter structure, commonly depicted as 
%
\[
    (p, d, q)
\]
where \textit{p} stands for the number of steps that a model needs to go back to the history data to predict the next value, that is, the level of autoregression (AR), \textit{ q} denotes the number of moving averages required (MA) and captures the short-term random fluctuations in the data, and \textit{d} defines the differencing level required to achieve stationarity in the trend observed within the data and therefore be able to make predictions~\cite{shumway2000time}. Models must be adjusted to these factors to reach stationarity in trained data. We provide a detailed description of how parameter tuning is undergone with ARIMA-based models in the appendix.

Similarly, models that capture seasonal patterns, such as SARIMA, are presented with an extended set of parameters
\[
    (p, d, q)(P, D, Q, m).
\]
When the model is adapted to capture the seasonal patterns of the data in models such as SARIMAX and SARIMA+LM, the parameters \textit{P, D, Q} are also calculated. These capture the impact of seasonality on the model parameters explained above, which the model adjusts based on the specified \textit{m} cycle level in the data. In our study, we consider \textit{monthly} and \textit{biweekly} cycles as previously defined. Thus, the adjusted models provide an additional layer in the model parameter tuning where the seasonality component is treated. 

In this study, the model parameters presented are tuned when the TSA models are fitted with the training data. To determine the optimal combination of the model parameters, we conducted an iterative process in which we searched for the best combination of model parameters for each of the project's time series data. Thus, for different combinations of the values of the parameters $p$ and $q$, we use the augmented Dickey-Fuller test~\cite{fuller2009introduction} to determine the optimal value \textit{d} of the diffencing model parameter. Similarly, when the seasonality effect is adjusted, the Canova-Hansen test~\cite{canova1995seasonal} is used to determine the optimal order of seasonal differentiation $D$, along with different combinations of $P$ and $Q$. To execute this iterative process, we adopt the \textit{Auto-arima} algorithm~\cite{hyndman2008automatic, wang2006characteristic} which defines the potential range of parameter values based on the fitted data. We provide a detailed description of the algorithm functioning in the appendix.

Based on the temporal patterns of the SQALE index, the different combinations of models computed for the defined parameters are evaluated using the Akaike Information Criterion~\cite{akaike1974new} (AIC) and the Bayesian Information Criterion~\cite{mcquarrie1998regression} (BIC) following the latest in statistical modelling~\cite{agresti2015foundations}. We provide a theoretical description of the model parameters presented in the appendix. Moreover, we provide the logic implemented in the shared online package (see Section~\ref{sec:Replicability}).

\subsubsection{Setting the stage for Machine Learning models (RQ$_2$ - RQ$_4$)}
\label{sec:ML}
\ReviewerA{In RQ$_2$ and RQ$_4$, we compare the prediction performance of TSA approaches with that of ML algorithms. Therefore, we also investigate the ability of ML models to predict code TD by applying a collection of linear and nonlinear ML algorithms. To run the comparison in equal conditions with all the models considered in this study, the same time series data used to train and test the TSA methods are also used to train and test the selected ML algorithms. Further descriptions of the training and performance evaluation of the models studied can be found in Section~\ref{sec:performanceval}.}

The ML methods used for the comparison are selected to provide a wide predictive perspective, as each of them textualizes different aspects of the data. Likewise, the chosen ML algorithms have been extensively used in the recent literature for their ability to predict software quality characteristics, such as code TD~\cite{Zozas2023, tsoukalas2021machine, Tsoukalas2020, Mathioudaki2021, Aversano2022}. 

The included ML models assuming \textbf{\textit{linearity} (L)} in the data are the following.
\begin{itemize} [leftmargin=*]
    \item \textbf{Multiple Linear Regression (MLR)} is a powerful statistical model commonly used to analyze the relationship between a continuous dependent variable and a set of independent explanatory variables. Through this relationship, the model can provide insight into how changes in the independent variables affect the value of the response variable while maintaining the linearity assumption~\cite{freund2006regression}.
    \item \textbf{Stochastic Gradient Descent (SGD)} algorithm is based on an iterative decrease in prediction loss by changing the model parameters. 
    Specifically, for each iteration, the SGD algorithm randomly selects a training set to bring randomness to the optimization problem, and the algorithm continues to optimize the result until it converges~\cite{bordes2009sgd}. 
    \item \textbf{Lasso regression (L1)} is a linear regression technique specialized in high-dimensional explanatory variable sets. This technique performs an optimization problem where model-independent variables are penalized if they are not important to the model. Thus, through model simplification, L1 achieves linear regression with the most relevant explanatory variables~\cite{tibshirani1996regression}. 
    \item \textbf{Ridge regression (L2)} is another linear regression technique that, through a regularization term, avoids overfitting the model. The L2 technique brings a penalization for the given model parameters that denote extreme values, and thus stabilizes the model, which results in advantages for applications where the number of model-independent variables is high~\cite{hoerl1970ridge}.   
\end{itemize}
The selected ML models assuming \textbf{\textit{nonlinearity} (NL)} are:
\begin{itemize} [leftmargin=*]   
    \item \textbf{Support Vector Machine (SVM)} is a well-established ML algorithm used for classification and regression tasks. The regression model aims to find the most optimal hyperplane that considers the characteristics of the existing explanatory variables to explain the values of the output response variable. For this, the model performs a minimization process of the deviation between the real value and the predicted value~\cite{de2003machine}. We considered using the so-called Gaussian kernel or Radial Basis Function as the function to build the model to get the perspective of a non-linear approach. 
    \item \textbf{Extreme Gradient Boost (XGB)} is an ML algorithm that is part of the family of gradient boost methods. Based on the ensemble learning methodology, the XGB algorithm employs multiple decision trees in a greedy format, where each of the subsequent trees corrects or refines the wrong results from the previous tree. Through this technique, the algorithm defines weights for the included independent variables that require a higher emphasis~\cite{chen2016xgboost}.
    \item \textbf{Random Forest regression (RF)} also belongs to the ensemble learning family and is built based on the ensemble of decision trees. Based on random sampling, the RF algorithm performs multiple decision trees and the output is the aggregated value from the results obtained in the trees performed~\cite{breiman2001random}.     
\end{itemize}

\subsubsection{Performance metrics (RQ$_1$ - RQ$_5$)} 
\label{sec:performancemet}

We used three performance metrics to compare and evaluate the prediction approaches and models.

The first performance metric is the \textit{Mean Absolute Percentage Error (MAPE)}. MAPE is a statistical metric commonly used to measure prediction precision. Quantifies the magnitude of errors between the predicted values and the actual observed values by calculating the mean of the absolute value of the prediction error. Its formula is the following.
    \begin{equation}
        MAPE = \frac{100}{n} \sum_{i=1}^n\frac{\mid(Y_i - \hat{Y_i})^2\mid}{Y_i}
     \end{equation}
where $n$ is the number of observations, $Y_i$ is the actual observed value, and $\hat{Y_i}$ is the i-th predicted value. As shown by the formula, MAPE is represented as a percentage metric, where a smaller percentage of error means better predictive performance, and the opposite is true with a higher percentage.

MAPE has limitations when the actual observations are small or close to zero, which introduces bias into the model training. Therefore, we selected alternative accuracy measurement statistics to cover this issue. One of them, and the second metric to present is the \textit{Mean Absolute Error (MAE)}, which is characterized by measuring the average magnitude from the absolute value of prediction errors. Its equation is
    \begin{equation}
         MAE = \frac{\sum_{i=1}^n \mid Y_i - \hat{Y_i}\mid}{n}
    \end{equation}
where the variables follow the same notation as the ones addressed for MAPE MAE expresses the prediction performance error in absolute value; therefore, a smaller result signifies better predictive performance, while higher results represent poorer performance.

We adopt as the third performance metric \textit{ root mean squared error} (RMSE) which captures the error value in the same value unit as the variable being predicted. Its formula is 
\begin{equation}
        RMSE = \sqrt{\frac{1}{n} \sum_{i=1}^n (Y_i - \hat{Y_i})^2}
     \end{equation}
where the variables follow the same notation as the ones explained previously.  Although the definitions have value
unit differences, a low error value indicates a high predictive performance in RMSE, and similarly, a high error
value depicts a poor predictive performance. We provide further results on AIC and BIC model selection criteria in the online Appendix (see Section~\ref{sec:Replicability}).

\subsubsection{Performance evaluation (RQ$_1$ - RQ$_5$)} 
\label{sec:performanceval}

\ReviewerA{To evaluate the performance of the confronted models, we trained and tested the TSA and ML models on the two time-series datasets generated during the data serialization in Section~\ref{sec:preprocessing_timeseries}. We adopted the Walk-Forward Train-Test validation technique~\cite{stone1974cross} for the performance evaluation of all models considered in this study. This technique was chosen because it respects the temporal nature of the data and does not rely on randomness in the order of the observations. This helped to produce a fair comparison between the confronted methods, as especially ML models are not structurally designed to understand the temporal order existing in the fitted data.}

\begin{figure}
    \centering
    \includegraphics[width=0.95\columnwidth]{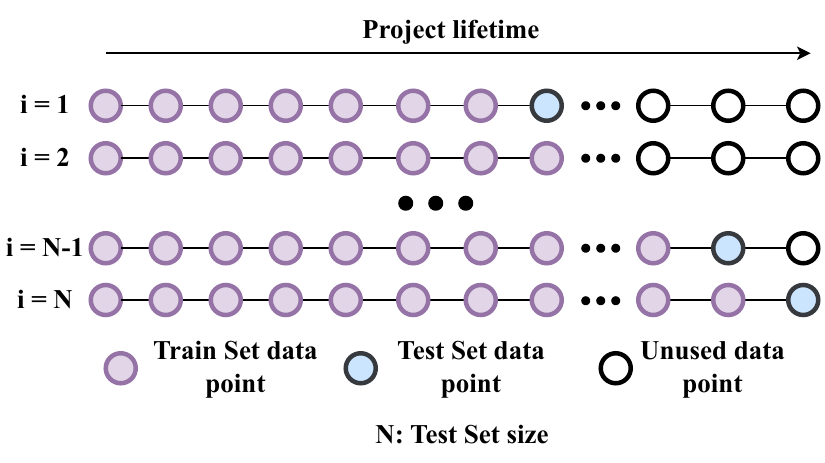}
    \caption{Walk-Forward Train-Testing approach. (\textit{i}: New observation.)}
    \label{fig:walkforward}
\end{figure}

The Walk-Forward Train-Test technique trains and tests a model greedily and iteratively. Each iteration trains a model, predicts the value of the next data point of the time series, and compares the prediction against the real value. The real value is consequently added to the train set, and the model is trained again with the newly extended train set data. This cycle continues until all test data points are predicted. In our study, the models start training with 80\% of the data as training data and then predict and test every new data point. Figure~\ref{fig:walkforward} shows a graphical representation of the approach, where \textit{i} denotes the new data point in each iteration. The described model training and testing phases performed in the performance evaluation stage of this study were implemented equally in each of the adopted models, with the same data testing set, for the sake of fairness in the predictive performance comparison.

\begin{ReviewerAEnv}

\subsubsection{Practitioners' forecasting perceived valuable (RQ$_6$)}
This section presents the data analysis of our work. Our survey includes closed and open questions. Therefore, we select different analysis methods for the two types of survey output. To analyze the responses to the closed questions, we initially employed descriptive statistics to gain a clearer understanding of the data. For ordinal and interval data, we focused on the mode and median to assess central tendency, while for nominal data, we calculated the distribution of participants' choices for each option.

Regarding the Likert scale question, Q$_{12}$, the possible value ranged from 0 (``not at all useful'') to 10 (``extremely useful''). We present both the values as they have been reported by the interviewees and also the interpretation of the Net Promoter Score (NPS) \cite{reichheld2003nps}.  NPS is a widely used metric for measuring customer loyalty and satisfaction. It classifies respondents into three categories based on their likelihood to recommend a product, service, or company:

\begin{itemize}
    \item \textbf{Promoters (9-10):} Highly satisfied and likely to advocate for the product.
    \item \textbf{Passives (7-8):} Moderately satisfied but not enthusiastic enough to promote it.
    \item \textbf{Detractors (0-6):} Unhappy customers who may discourage others from using the product.
\end{itemize}
The NPS is calculated as: $NPS = \text{Percentage of Promoters} - \text{Percentage of Detractors}$. Therefore, a higher NPS indicates stronger customer loyalty. Hence, we employ such a score to measure the practitioner's perceived value of our model.

Regarding open-ended questions, we employ qualitative data analysis techniques suggested by Strauss and Corbin \cite{straus1998techniques} and Seaman and Yuepu \cite{seaman2011measuring}. Qualitative analysis assists in addressing questions such as ``What is happening in this situation?'' when we aim to uncover how individuals understand their experiences and manage them over time in amidst evolving conditions \cite{rios2020practitioners}. 

We adopted an inductive approach to develop a new theory based on qualitative data. The open questions for RQ$_{6.1}$ and RQ$_{6.2}$ were manually coded as follows. Two authors independently coded responses to related questions adopting thematic analysis~\citep{straus1998techniques,esposito2024large}. We addressed disagreement through discussion, and codes were organized into a hierarchy of benefits and limitations until saturation was reached. 

\end{ReviewerAEnv}

\subsection{Model Execution (RQ$_{1}$ - RQ$_{5}$)} 
\label{sec:modelexec}

In this section, we textualize the description of the model execution process of the defined TSA models to facilitate practitioners' understanding of the results presented in Section~\ref{sec:Result} and encourage the use of the proposed models in the SE community. However, the prediction process described in Section~\ref{sec:performanceval} is followed similarly for all projects with each of the prediction models considered using different data sets. After the prediction process is completed, the results are visualized and, therefore, organized by aggregating the results provided by each model with all the collected projects for the two defined data sets (biweekly data and monthly data). The aggregation of the results per data set for each model combination is done by taking the average performance of the different projects in the respective data set as displayed in Table~\ref{tab:biweeklyPerformance} and Table~\ref{tab:monthlyPerformance} for instance.

\subsubsection{ARIMAX \& SARIMAX models (RQ$_1$ - RQ$_4$)}
\label{sec:sarimax_exec}

Within the context of the modeling process, both the ARIMAX and SARIMAX models are built similarly in their initial stages. The last one stands as an adjusted version of the former counterpart. Therefore, for each project, we built both an ARIMAX model and a SARIMAX model. \ReviewerB{We computed the exploratory analysis of the model variables to examine their distributional characteristics (see Table~\ref{tab:modelVariables} and Figure~\ref{fig:variables_boxplot}). All model variables denoted high skewness values, which was visually represented in the displayed distribution boxplots. Therefore, we applied the \textit{log-transform} technique on the independent variables of the model to standardize their distribution and thus reduce the complexity of the model variables during the learning process.} Variable transformation is a common preprocessing technique used in the field of predictive analytics due to its benefits in reducing the complexity of the model and increasing the prediction performance~\cite{flach2012machine}.  


\ReviewerA{The first step of the modeling process consisted of leveraging the \textit{Backward variable selection} technique~\cite{agresti2015foundations}. The stepwise backward procedure begins with fitting the model with the set of important independent variables initially considered. Subsequently, it sequentially assesses the model and removes variables from the model to train it back. In each sequence, it selects the variable whose removal improves the goodness of fit of the model. The process stops when any further removal leads to a poorer model fit, therefore selecting the set of independent variables that provides the most optimal model results. To assess the impact of each combination of variables accordingly and following the concepts described in Section~\ref{sec:parameterTuning}, we performed the model parameter tuning approach with the \textit{Auto-arima} algorithm to obtain the results of the criteria AIC and BIC. We provide a concise description of the parameter tuning process in appendix A.}

\ReviewerA{Figure~\ref{fig:decomposition} provides a graphical representation of the seasonal decomposition process performed for the SARIMAX model with the biweekly SQALE index data of the \textit{Apache httpcore} project to achieve the best model parameters  Described in a top-down structure, the first plot provides the observed progress of the SQALE index across time, the second and the third plots provide the existing trend and seasonality pattern, and the fourth one depicts the remaining residuals or \textit{noise} in the data after de-trending and de-serializing has been performed  The seasonal decomposition process is conducted every time a seasonally adjusted model is built  Hence, we provide the visualization of the seasonal decomposition for each of the considered projects in the provided online appendix (see Section~\ref{sec:Replicability}).}

\begin{figure}
    \centering
    \includegraphics[width=1\columnwidth, trim={10 0 0 0},clip]{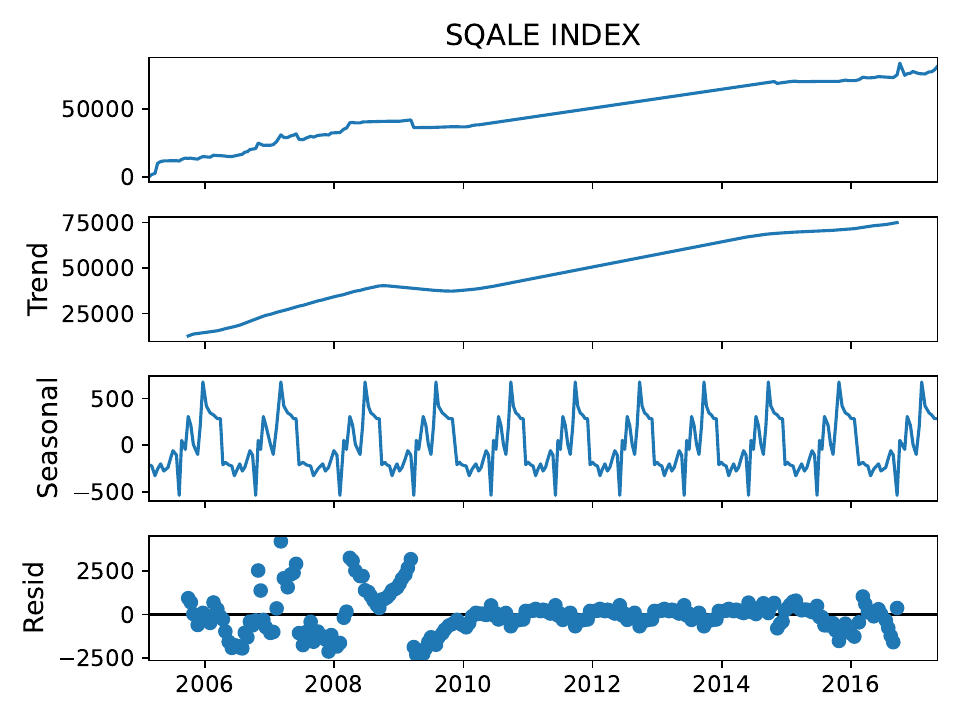}
    \caption{Seasonal decomposition of the SQALE index for project \textit{httpcore} with \textit{biweekly} data.}
    \label{fig:decomposition}
\end{figure}

\begin{figure}
    \centering
    \includegraphics[width=1\columnwidth, trim={10 10 30 30},clip]{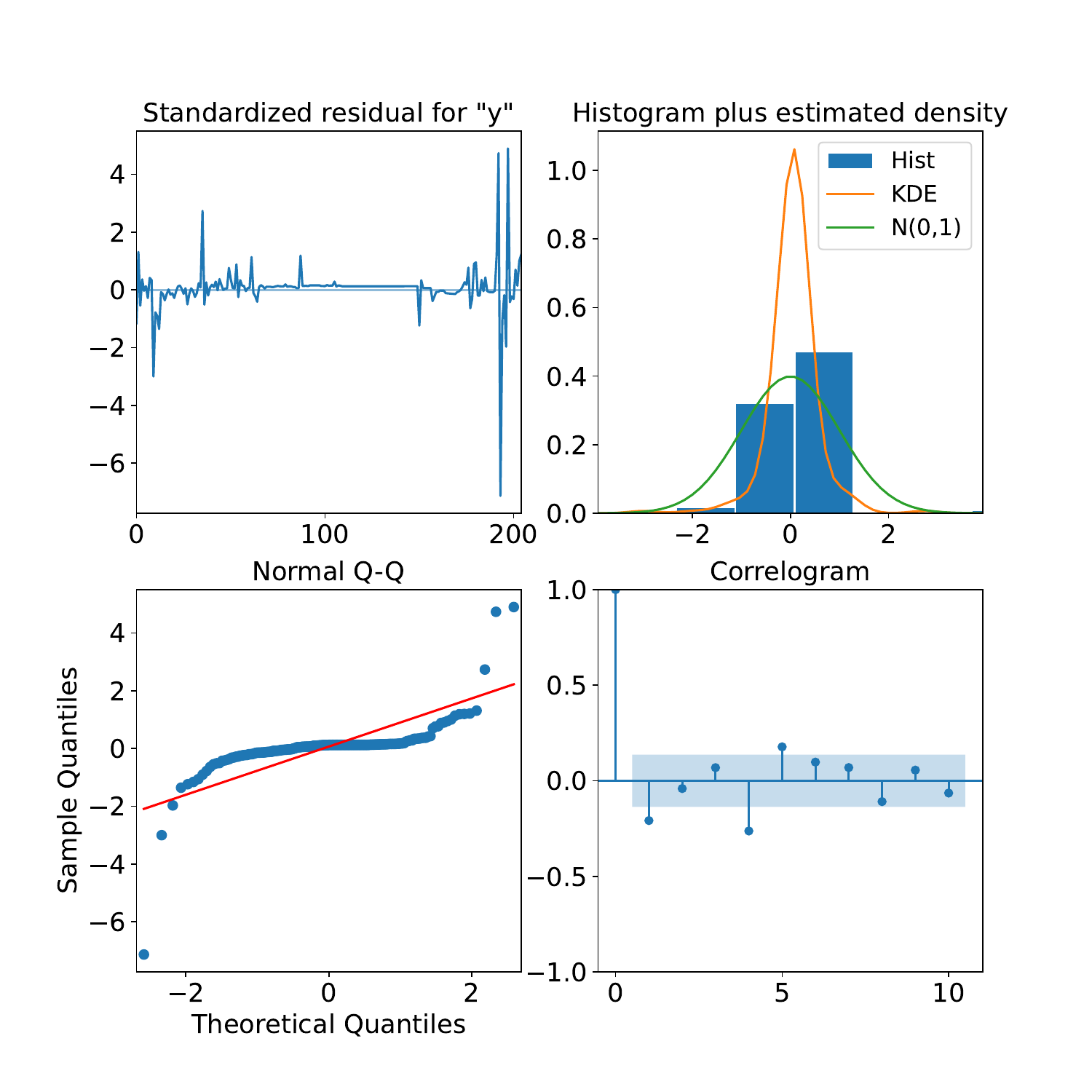}
    \caption{Final model diagnostics of the SARIMAX model built for project \textit{httpcore} with \textit{biweekly} data.}
    \label{fig:model-diagnostics}
\end{figure}

\ReviewerA{For each of the studied software projects, we performed the multivariate TSA model fitting following the explained stepwise process. After each process, the model would provide the best goodness-of-fit results for the fitted data, and normally distributed model residuals, therefore showing good model quality. We provide a graphical representation of this stage in Figure~\ref{fig:model-diagnostics}, where the model diagnostics are presented for the same example project as in Figure~\ref{fig:decomposition}.}

\ReviewerA{The standardized residuals help to identify the existing anomalies or obvious patterns if the former ones are not centered around zero, therefore meaning that some patterns remain uncaptured by the model. The second plot provides the histogram of the model residuals, where in comparison with the standard normal distribution the kernel density estimate seems to be approximately normally distributed. In the third plot, the model residuals are compared against the Q-Q plot standards, showing a deviated distribution of the residuals from normality. The correlogram or Auto Correlation Function (ACF) plot shows the residuals autocorrelations, which should not be present after the model is fitted. Hence, they should not be exaggerated outside the marked confidence interval.}

\ReviewerA{Finally, we tested the resulting models through the performance evaluation process explained in Section~\ref{sec:performanceval}, and similarly, we calculated the results from the performance metrics as described in Section~\ref{sec:performancemet}.}

\subsubsection{ARIMA+LM and SARIMA + LM models (RQ$_1$ - RQ$_4$)}
\label{sec:sarima_exec}

Similarly to the previous section, the ARIMA+LM and SARIMA+LM approaches share the same model-fitting logic, with the difference being the adjustment of the seasonality effect. Therefore, for the sake of brevity, if we consider, for instance, the SARIMA+LM approach, it fits a SARIMA univariate model for each of the independent variables, thus predicting the future values of each variable independently and only based on the past historical observations. The LM model is then fitted with the predicted values of the independent variables to predict future values of the SQALE index (see Section~\ref{sec:ARIMA}). Then, we perform the parameter turning process for each of the models to achieve the best combination of model parameters to be used for prediction, as shown in~\ref{sec:parameterTuning}. We then predicted the future values of each of the independent variables through their trained SARIMA models, respectively, following the performance evaluation methodology described in Section~\ref{sec:performanceval}. With the future values of the independent variables already predicted, we trained the LM model with the training set of the historical SQALE index values along with the historical values of the independent variables. Thus, during the testing stage, we used the independently predicted values of the independent variables to predict future SQALE index values. The performance evaluation of the linear regression model was calculated in the same format as previously done in Section~\ref{sec:sarimax_exec}.

\ReviewerA{\subsubsection{Performing long-term forecasting with TSA methods (RQ$_5$)}}
\label{sec:long_term_analysis}

\ReviewerA{Following the motivation stated in Section~\ref{sec:ESDesign} for RQ$_5$, this section presents the process executed to perform long-term forecasting with TSA methods. For that, we selected the TSA methods that show better performance from the results obtained in the model comparison analysis performed to answer RQs 1 to 4. To provide the models considered with a longer forecast horizon, we selected an initial data split of 70\% for training and 30\% to test the data set in which each model presented their best performance accordingly (biweekly or monthly data). We used the best model parameter and variable configurations (see Table~\ref{tab:parametersummary} obtained from the previous analysis stages to provide the forecasts with the best detected model settings.}  

\ReviewerA{To put ourselves in the shoes of system engineers, the MAPE metric was chosen to quantify the prediction error obtained in each of the forecasted time-steps, given its easier interpretability. Since we are interested in evaluating the long-term forecast performance of the models considered, we did not retrain the models with each subsequent time point tested as described in Section~\ref{sec:performanceval}.}

\ReviewerA{Moreover, since we are calculating the performance level of TSA models across multiple software projects, we aggregated the MAPE results in each time-point, and computed mean, variance, maximum, minimum and median statistics from the drawn distributions. Since the task of assessing ``once and for all'' what the right precision would be due to the ``magic'' nature of thresholds, we also extend the interpretation of our results in the discussion section.}

\section{Results}
\label{sec:Result}

\begin{figure*}
    \centering
    \includegraphics[width=1\textwidth]{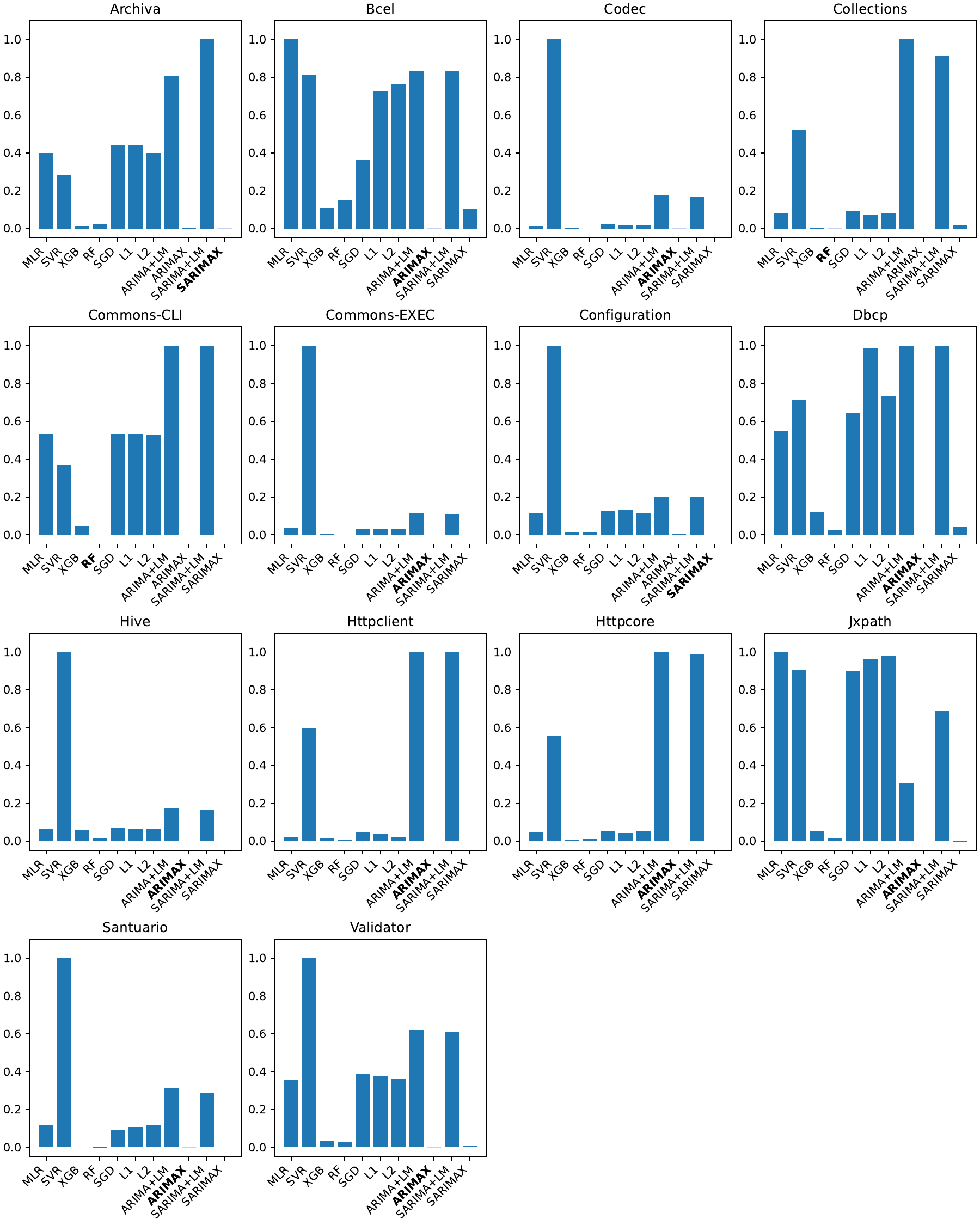}
    \caption{MAE prediction performance results using the biweekly data for the 14 analyzed projects.}
    \label{fig:global-mae-results}
\end{figure*}

In this section, we report the results obtained by answering our RQs. The data set included original commit data for 31 OS Java projects, and serialization and posterior linear interpolation created data serialized in biweekly and monthly time series. We successfully processed 14 software projects with all the models used in this study. We encountered convergence issues with linear algebra while fitting the TSA models. We report this issue in the model fitting stage as a threat to the validity of this study, which we formally state in Section~\ref{sec:Threats} and provide further details on it later in this section. Therefore, since we encountered this issue in the execution of different models across the collected projects and for the sake of fairness in the models' results, we excluded projects that could not provide results for all the models considered in the study. To answer our RQs, we present the results of the analysis performed as the average score of the outputs obtained with each prediction model for the defined prediction performance metrics. Similarly, we perform the aggregation step separately for the biweekly and monthly data results. Figure~\ref{fig:global-mae-results} illustrates the MAE results for the models executed in the projects analyzed. We provide all results and tables before aggregation in the shared online package (see Section~\ref{sec:Replicability}).

\subsection{Descriptive Analysis} 
\label{sec:descriptivestats}

\begin{ReviewerAEnv}
We now present the descriptive statistics of the subjects initially considered before building our prediction models. Table~\ref{tab:descriptiveStats} (see the appendix) presents the statistical descriptive characteristics of the projects based on the raw data set, as well as the generated biweekly and monthly time series data. According to Table~\ref{tab:descriptiveStats}, we identified two opposite trends in the number of SQ analysis executions carried out by the study projects. Certain projects exhibited a decrease in the number of SQ analysis executions when the original observations from the raw data set were serialized into time series data (e.g., \textit{Felix} in Table~\ref{tab:descriptiveStats}). This finding indicated that these projects registered brief periods of time in which an elevated number of SQ analysis executions were performed, which subsequently decreased over time, resulting in the absence of SQ analysis executions during extended periods. Therefore, while generating the time series data, multiple executions of the SQ analysis that occurred within a biweekly or monthly time observation were converted into a single observation, as explained in Section~\ref{sec:preprocessing_timeseries}. However, some other projects presented the opposite scenario, as their number of observations increased when the time series data were generated (e.g. \textit{ fileupload} in Table~\ref{tab:descriptiveStats}). This finding indicated that these projects registered a limited number of SQ analysis executions during the time frame used to collect data for the Technical Debt dataset.  Consequently, this type of project required data interpolation for some of the time-series observations due to the absence of real SQ analysis executions.

At this stage of the analysis, we realized that, depending on the frequency of the SQ analysis executions, the quality of the data would vary. The variations presented suggested that performing static analysis with tools like SQ in a fixed frequency could result in concise, authentic, and well-monitored time series data on the state of the quality of the code base of projects. In contrast, less frequent or randomly executed SQ analyzes provide insight into the code base throughout time but leave notable periods with missing data, thus requiring the action of data interpolation to perform TSA. Moreover, due to missing data for some of the projects, the algorithms of the adopted TSA models failed to converge during the model fitting stage described in Section~\ref{sec:ARIMA}. Due to the existing complexity in the data, we encountered several linear algebra decomposition issues~\cite{golub2013matrix} involved in the parameter estimation process while fitting the TSA models. As mentioned earlier, we leveraged preprocessing the set of independent variables within the model training through the backward variable selection criteria and standardization. However, Table~\ref{tab:resultingprojects} (see the Appendix) reports the cases in which the adopted models reported linear algebra decomposition issues, which impeded obtaining results for all the models in each project that reported issues, respectively.
\end{ReviewerAEnv}

\begin{figure*}
        \centering
        \begin{subfigure}[t]{0.3\textwidth}
            \centering
    \includegraphics[width=\columnwidth]{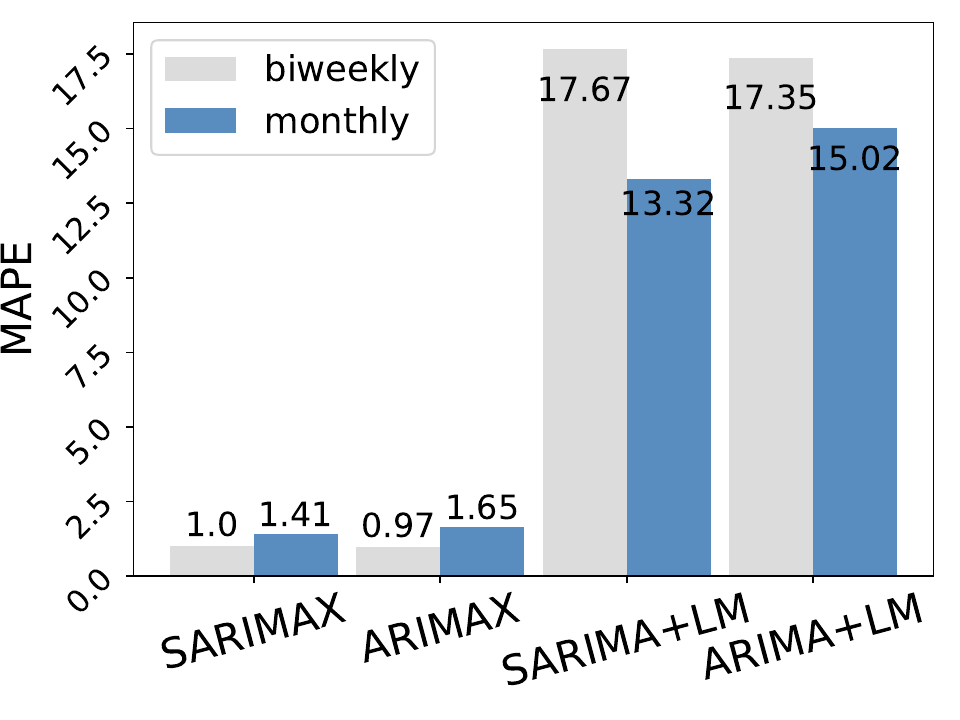}
            \caption{\textbf{MAPE}}
            \label{fig:mape}
        \end{subfigure}
        \hfill
        \begin{subfigure}[t]{0.3\textwidth}
            \centering
    \includegraphics[width=\columnwidth]{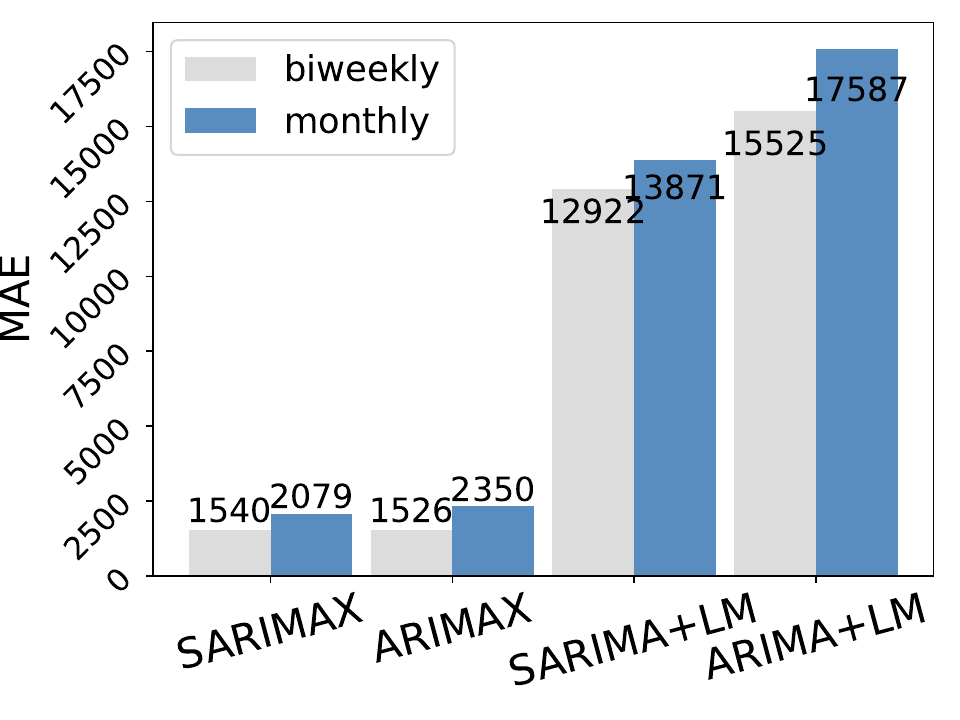}
    \caption{\textbf{MAE}}
    \label{fig:mae}
    \end{subfigure}
        \hfill
        \begin{subfigure}[t]{0.3\textwidth}
            \centering
    \includegraphics[width=\columnwidth]{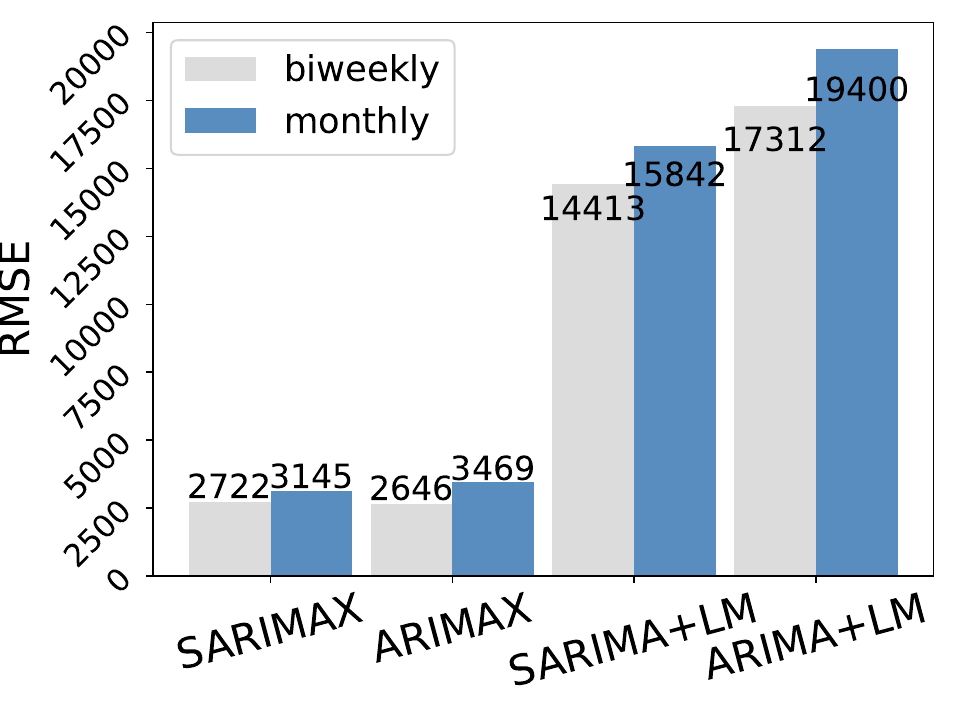}
    \caption{\textbf{RMSE}}
    \label{fig:rmse} 
        \end{subfigure}
        \caption{Prediction performance (MAPE, MAE, and RMSE) for the considered TSA models.}
    \label{fig:rq1_results}
\end{figure*}

\subsection{On the performance of Code TD prediction with existing time-dependent approaches (RQ$_1$)}

We conducted the prediction analysis described in Section~\ref{sec:modelexec} for each of the 31 projects in our data set. Figure~\ref{fig:rq1_results} shows the unified graphical illustration of the predictive performance results of the performance metrics adopted. The displayed results demonstrate a clear superiority of ARIMAX compared to ARIMA+LM with biweekly and monthly serialized time series data. Taking into account the results for MAE and RMSE, the differences in prediction error between ARIMAX and ARIMA + LM are 14,000 and 14,666 within the biweekly data and 15,237 and 15,931 within the monthly data, thus denoting a remarkable difference between both approaches. Hence, we can affirm that \textbf{given the existing TSA models for code TD prediction, the ARIMAX model overcomes the predictive performance of the ARIMA+LM model}. 
Finally, both models performed better with biweekly training data, as shown by MAE and RMSE. However, MAPE indicated an improved predictive performance for the ARIMA+LM model with monthly training data.

\subsection{On the comparison of time-dependent approaches and ML models for Code TD prediction (RQ$_2$)}

We compared the ARIMAX and ARIMA+LM models against the ML algorithms commonly adopted in the field. Tables~\ref{tab:biweeklyPerformance} and~\ref{tab:monthlyPerformance} present the prediction performance results for the biweekly and monthly data, respectively. More specifically, \textbf{the tables provide the average results calculated from the prediction results obtained for the 14 projects}. The results of the individual projects are included in the replication package. It should be noted that, in both cases, using biweekly and monthly data, \textbf{ARIMAX model demonstrated a clear superiority over the rest of the models}, thus providing a more accurate prediction outcome for future code TD observations. Results such as 1,525.96 for MAE and 2,646.20 for RMSE within the biweekly results, compared to those presented by the following model after the ranking of results, RF, denote a quantified improvement of 717 and 1,232 in the described prediction errors accordingly. In addition, RF proved to be the second-best prediction model. As described previously, RF assumes nonlinear relationships between variables. Consequently, despite the indications that TSA models such as ARIMAX may be appropriate for predicting code TD, the potential for nonlinear models to capture intricate relationships cannot be discounted. Therefore, we observe that such a finding is especially relevant given the possibility of incorporating additional data and model predictors. Except for the special case of SVM, non-linear models consistently outperform linear models across all calculated metrics.

\begin{table}
    \caption{Comparison of the predictive performance of TSA models against ML models for \textit{biweekly} data. (\textit{L}: Linear, \textit{ML}: Non-linear)}
    \label{tab:biweeklyPerformance}
    \footnotesize
    \centering
    \begin{tabular}{@{}l|rrr@{}}
         \textbf{Approach} & \textbf{MAPE (\%)} & \textbf{MAE} & \textbf{RMSE}\\
         \hline
         \textit{SARIMAX} & 1 & 1,539.92 & 2,721.95\\
         \textit{SARIMA + LM} & 17.67 & 12,921.79 & 14,412.90\\
         \colorbox{gray!30}{\textbf{\textit{ARIMAX}}} & \colorbox{gray!30}{\textbf{0.97}} & \colorbox{gray!30}{\textbf{1,525.96}} & \colorbox{gray!30}{\textbf{2,646.20}}\\
         \textit{ARIMA + LM} & 17.35 & 15,525.21 & 17,312.14\\
         \textit{MLR$_{(L)}$} & 4.85 & 4,895.50 & 6,759.43\\
         \textit{SVM$_{(NL)}$} & 23.87 & 38,929.00 & 39,904.36\\
         \textit{XGB$_{(NL)}$} & 1.72 & 3,006.66 & 4,686.55\\
         \textit{RF$_{(NL)}$} & 1.44 & 2,242.57 & 3,878.84\\
         \textit{SGD$_{(L)}$} & 5.04 & 5,494.12 & 7,275.05\\
         \textit{L1$_{(L)}$} & 5.18 & 5,431.16  & 7,112.04\\
         \textit{L2$_{(L)}$} & 4.86 & 4,898.60 & 6,709.81\\
         \hline
    \end{tabular}
\end{table}

\begin{table}
    \caption{Comparison of the predictive performance of TSA models against ML models for \textit{monthly} data. (\textit{L}: Linear, \textit{ML}: Non-linear)}
    \label{tab:monthlyPerformance}
    \footnotesize
    \centering
    \begin{tabular}{@{}l|rrr@{}}
         \textbf{Approach} & \textbf{MAPE (\%)} & \textbf{MAE} & \textbf{RMSE}\\
         \hline
         \colorbox{gray!30}{\textbf{\textit{SARIMAX}}} & \colorbox{gray!30}{\textbf{1.41}} & \colorbox{gray!30}{\textbf{2,078.64}} & \colorbox{gray!30}{\textbf{3,145.27}}\\
         \textit{SARIMA + LM} & 13.32 & 13,871.48 & 15,842.22\\
         \textit{ARIMAX} & 1.65 & 2,349.81 & 3,469.03\\
         \textit{ARIMA + LM} & 15.02 & 17,586.97 & 19,400.41\\
         \textit{MLR$_{(L)}$} & 4.62 & 3,978.65 & 5,567.30\\
         \textit{SVM$_{(NL)}$} & 23.12 & 27,822.99 & 28,620.02\\
         \textit{XGB$_{(NL)}$} & 5.80 & 4,600.92 & 5,950.18\\
         \textit{RF$_{(NL)}$} & 1.90 & 2,572.85 & 4,202.80\\
         \textit{SGD$_{(L)}$} & 4.89 & 4,478.40 & 5,956.63\\
         \textit{L1$_{(L)}$} & 5.09 & 4,609.15 & 6,067.31\\
         \textit{L2$_{(L)}$} & 4.50 & 3,706.95 & 5,192.83\\
         \hline
    \end{tabular}
\end{table}

\subsection{On the impact of seasonality on Code TD prediction in time-dependent approaches (RQ$_3$)}
\label{sec:seasonality_impact}

Following the set of defined research questions, we wanted to analyze the impact of addressing the seasonality effect on the prediction performance of TSA models through our RQ$_3$. Thus, we performed the prediction analysis described in Section~\ref{sec:modelexec} with SARIMAX and SARIMA+LM models. The process was similar to RQ$_1$ and used the exact data for this RQ. Figure~\ref{fig:rq1_results} shows the unified graphical representation of the predictive performance results of the performance metrics adopted. To avoid overwhelming the reader with too many plots, we displayed the results for the seasonally adjusted models and their former counterparts, ARIMAX and ARIMA+LM. The results displayed demonstrate a systematic difference in prediction performance between the multivariate single model approach, i.e., ARIMAX and SARIMAX, compared to that of the multivariate combined model approach, i.e., ARIMA+LM and SARIMA+LM. Therefore, we can affirm that \textbf{ the superiority in the prediction performance demonstrated on ARIMAX compared to ARIMA+LM is maintained with SARIMAX and SARIMA+LM when adjusting the models to capture seasonality}. Figure~\ref{fig:rq1_results} demonstrates the clear superiority of SARIMAX over SARIMA + LM given serialized time series data biweekly and monthly. Hence, we can affirm that given the adjustment of the existing TSA models for code TD prediction, \textbf{the SARIMAX model outperforms the predictive performance of the SARIMA+LM model}. Specifically, for MAE and RMSE, the prediction error difference between SARIMAX and SARIMA+LM is 11,382 and 11,691 for biweekly data and 11,792 and 12,697 for monthly data, respectively. These values highlight a significant gap between the two approaches. 

Hence, to answer RQ$_3$, the results displayed provide slightly similar results between the SARIMAX and ARIMAX models. The results for the biweekly data provided better prediction performance for ARIMAX than SARIMAX, while the opposite happened for the monthly data. MAPE results provided a worse prediction performance for SARIMA+LM than for ARIMA+LM when using biweekly data. However, SARIMA+LM overtook ARIMA+LM for the MAPE results when using monthly data and in the other prediction performance evaluation metric results. Therefore, we can affirm that \textbf{adjusting the TSA models to capture the seasonality effects of the data improves their prediction performance to some extent}.

\subsection{On the comparison of seasonally adjusted time-dependent models and ML models for Code TD prediction (RQ$_4$)}
\label{sec:TSAvsML}

We compared the previously confronted SARIMAX and SARIMA+LM models against the adopted ML algorithms commonly used in the field. Tables~\ref{tab:biweeklyPerformance} and~\ref{tab:monthlyPerformance} present the prediction performance results for the SARIMAX and \mbox{SARIMA + LM} models compared to the adopted ML models. As expected from the results obtained for RQ$_3$, where SARIMAX provided similar prediction results to ARIMAX when using biweekly data and overtook ARIMAX when using monthly data, in this case, \textbf{the SARIMAX model also demonstrated a clear superiority among the rest of the compared models}, thus providing a more accurate prediction outcome for future code TD observations.
It should be noted that capturing the seasonality effect did not change the key results observed for RQ$_{2}$. SARIMAX did not improve the results of the ARIMAX model using biweekly data, and the improvement experimentally compared to ARIMAX, with low monthly results. In addition, RF was shown to be the second-best prediction model. Therefore, even if the results suggest the suitability for linear models, nonlinear models may also offer the potential to capture more complex relationships. Looking at the rest of the ML models and excluding the special case of the SVM model, non-linear models present more accurate results than linear models in all the calculated metrics.

\subsection{\ReviewerA{On the long-term forecasting accuracy with TSA models (RQ$_5$)}}

\ReviewerA{Considering the single-step prediction performance presented in Section~\ref{sec:seasonality_impact} and Section~\ref{sec:TSAvsML}, the results of this study provided practitioners with two main TSA models that performed well in predicting future code TD one month ahead and two weeks ahead, we extend the result of the resulting best models, ARIMAX and SARIMAX, by exploring their long-term forecast performance.}

\begin{figure*}[htb]
    \centering
    \includegraphics[width=0.9\linewidth]{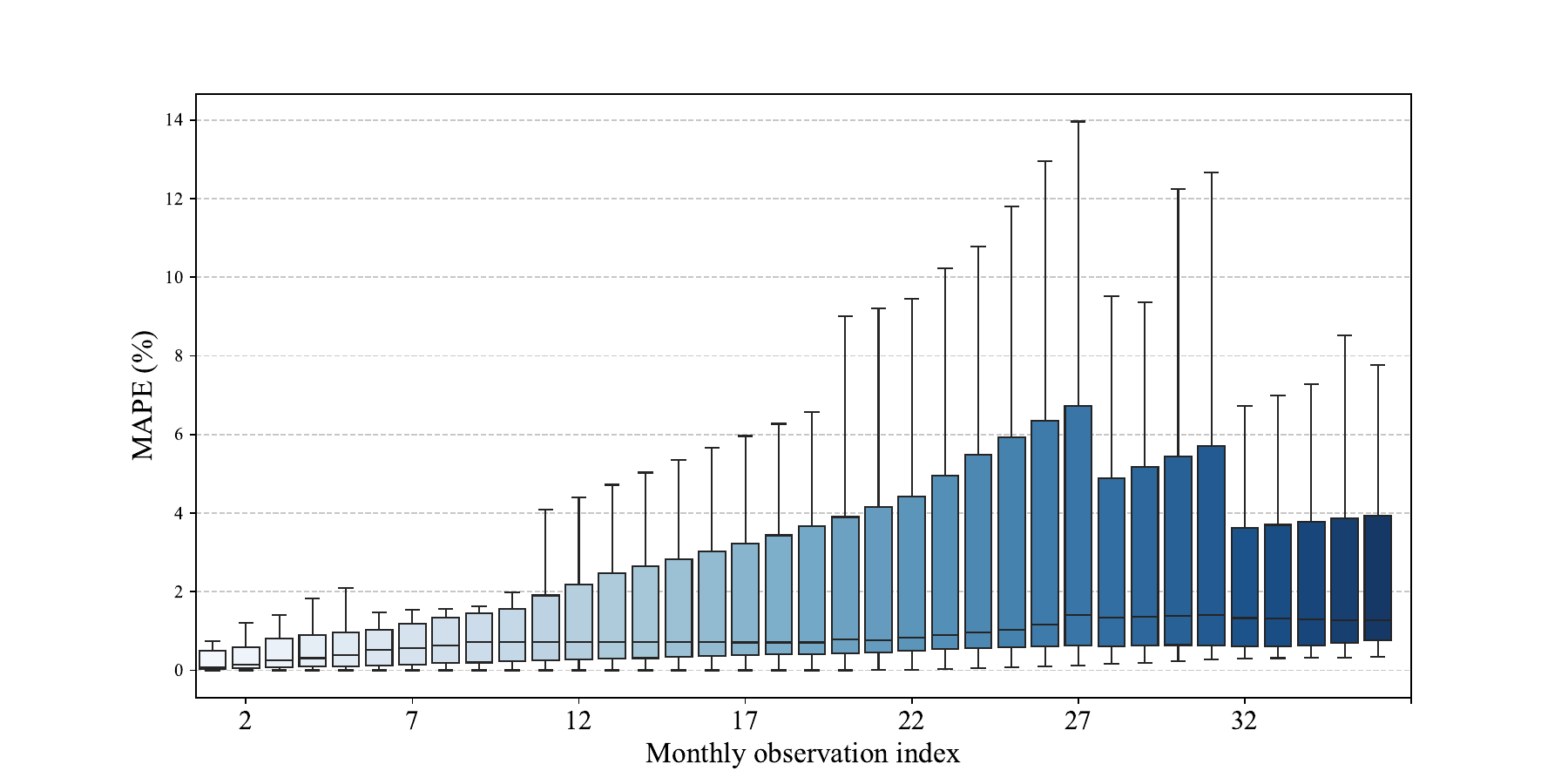}
    \caption{SARIMAX 36-month MAPE results with monthly data.}
    \label{fig:long_term_sarimax}
\end{figure*}

\begin{figure*}[htb]
    \centering
    \includegraphics[width=0.9\linewidth]{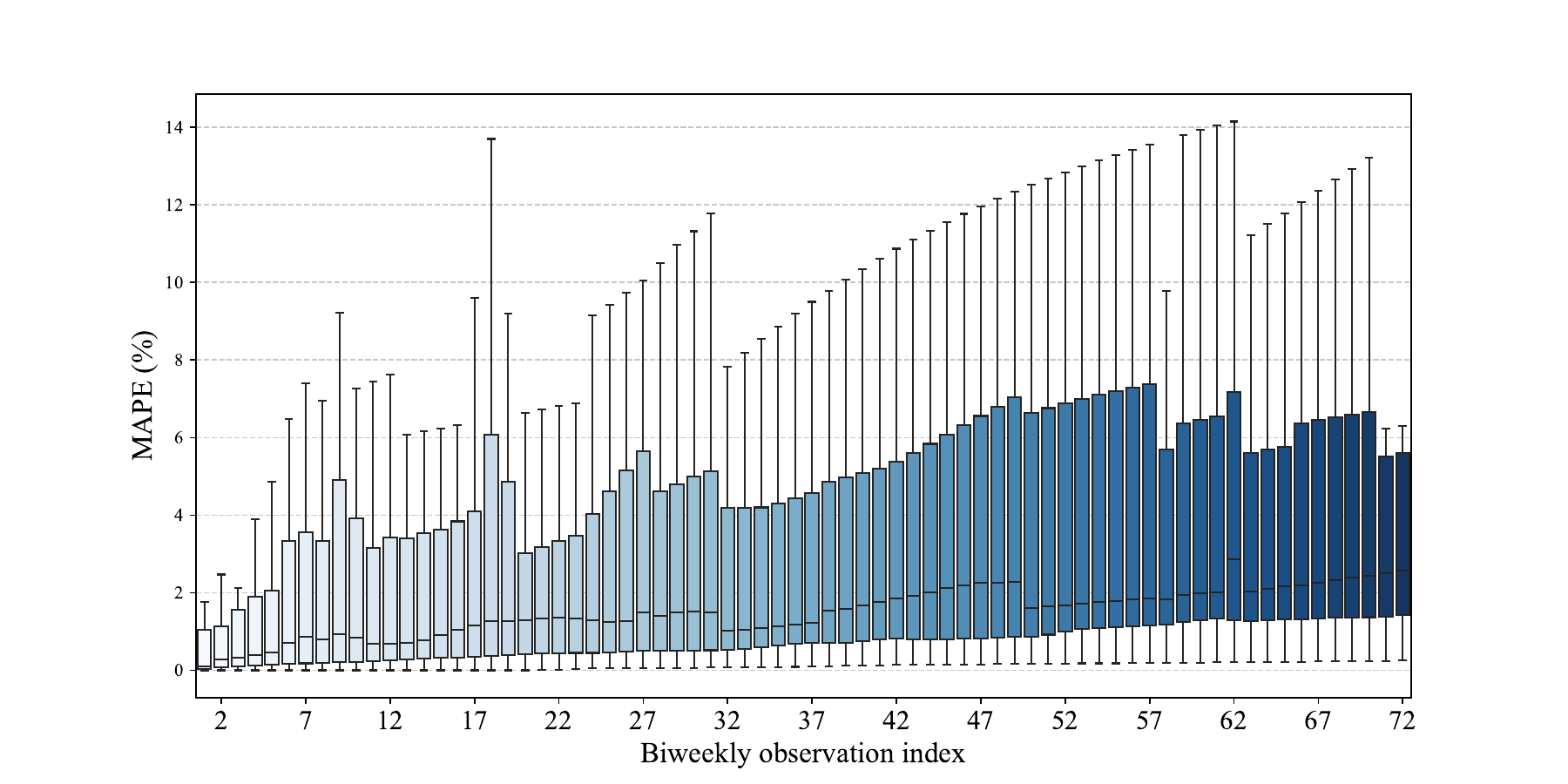}
    \caption{ARIMAX 72-biweekly (36 months) MAPE results with biweekly data.}
    \label{fig:long_term_arimax}
\end{figure*}

\ReviewerA{Following the data analysis process described in Section~\ref{sec:long_term_analysis} the ARIMAX model was trained with the biweekly dataset, and the SARIMAX model was trained with the monthly data set. These models were selected based on their demonstrated superior performance among the models compared in the study, and each serialized data set was used accordingly. For each project time series data, the models were trained with 70\% of the series, the remaining 30\% being used to assess the performance of the forecasts without retraining the models. Figure~\ref{fig:long_term_sarimax} illustrates the results of the long-term prediction of the SQALE index for the SARIMAX model visualized through the quantified MAPE errors, considering the monthly time series data for the 14 projects. Similarly, Figure~\ref{fig:long_term_arimax} presents the MAPE errors resulting from the long-term SQALE index forecasting for the ARIMAX model, considering the biweekly time series data in the equivalent number of projects.}

\begin{ReviewerAEnv}
It is important to note that each time point in the X-axis of both figures represents a new time period; for instance, time point 1 in the first figure represents the forecast for the future next month from the current time point. We displayed a box plot for each future time observation with the distribution of the resulting MAPE values from forecasts made on the resulting projects. Within the displayed boxplots we can find the median value of the obtained MAPE result, as well as the maximum and minimum values. Given the nature of the data, code TD, we considered limiting the total forecasting window length to 3 years, counted as 36 months and 72 biweekly periods.

Also, given that some projects had longer history data than others (see Table~\ref{tab:descriptiveStats} in the appendix), we could observe sudden drops in the magnitude of the boxplot in some time points, which depicted the end of the historical data of one of the resulting projects. However, this fact did not affect the evidence of a constant low-variance increasing cross-project code TD trend, which suggested a clear decline in the forecasting performance of the trained models, yet balanced over time. Furthermore, following the reported analysis design, we also calculated summary statistics from the aggregated MAPE results to quantitatively complement the displayed forecast trends. Table~\ref{tab:biweeklyforecaststats} and Table~\ref{tab:monthlyforecaststats} in the appendix present the results of the summary statistics.

The resulting values of the forecasts performed reaffirmed the key takeaways observed from the temporal trends previously displayed in Figure~\ref{fig:long_term_sarimax} and Figure~\ref{fig:long_term_arimax}. On the one hand, we observed a clear increase in the forecasting error quantified through the average MAPE values, a trend equally observed in the values for the median and the variation. On the other hand, we observed a relatively balanced trend in all the descriptive statistics mentioned. For example, interpreting the results from the perspective of a practitioner, the mean of the resulting aggregated monthly MAPE error from the SARIMAX model (see Table~\ref{tab:monthlyforecaststats}) increased only to 1.08\% for a forecast window of six months, 2.33\% for a forecast of twelve months, and 4.19\% for forecasts 36 months ahead of the current point. Regarding the variance of the results in the same table, its value was maintained below 5\% six months ahead of the current time point and almost reached less than 15\% of the variance being twelve months ahead. Looking at the forecasting results from the ARIMAX model with biweekly time series data, the increase in statistics such as the mean and the median remained balanced, although the variance of the distribution was already from the first forecast biweekly time point. However, the mean of the MAPE error values obtained remained at 3.15\% after 12 periods of biweekly (six months) and 4.56\% for 24 periods (twelve months). Even if the mean value oscillated over higher values in the subsequent biweekly periods, the final value after 72 periods (thirty-six months) from the current time point remained at 5.97\%.
\end{ReviewerAEnv}

\subsection{\ReviewerA{On the practitioners' forecasting perceived valuable  (RQ$_6$)}}
\begin{ReviewerAEnv}
Table \ref{tab:perceivedvalue} shows the score for the perceived value of our model. Using the Likert scale provided in the questionnaire and reported in Table \ref{tab:model_usefulness_scale}, according to Table \ref{tab:perceivedvalue}, 70\% of the participants perceived our model as useful to extremely useful, thus showing great appreciation for its possible effectiveness in the prediction of TD. 21\% of the interviewees deemed our model moderately useful and only 8\% deemed it very lowly useful.
More specifically,  Table \ref{tab:nps_distribution} presents the distribution of responses according to the Net Promoter Score (NPS) scale. The table categorizes the respondents into three groups: \textit{Promoters} (who rate the tool between 9 and 10), \textit{Passives} (who rate it between 7 and 8), and \textit{Detractors} (who rate it between 0 and 6). These categories reflect varying levels of enthusiasm and the likelihood of recommending the tool.

According to Table \ref{tab:nps_distribution}, the majority of respondents fall into the \textbf{Passives} category (47.83\%), indicating a generally satisfied user base that does not strongly advocate for the tool. The \textbf{Promoters} constitute 21.74\% of responses, representing those who are highly enthusiastic and are likely to recommend it. Conversely, \textbf{Detractors} account for 30.43\%, suggesting a notable portion of users who may be dissatisfied or unlikely to endorse the tool.  The computed NPS is 39.1, which falls into the ``\textbf{Neutral to Good}.'' This means that while there is a reasonable level of positive sentiment, there is still room for improvement in converting passive and detracting users into promoters.

Table \ref{tab:justifications_q12} also describes the justifications for these ratings. The strongest justification for positive ratings is that 17.39\% of the respondents identify the effectiveness the model could bring to the monitoring and treatment of TD. Another 13.04\% value its predictive aspect, citing its ability to prevent and control.
Some of the participants (4.35\%) liked its application in effort estimation, planning, and risk reduction, reaffirming the real-world advantages of our solution. In addition, none of the participants rejected the model as completely useless (0\%), validating the usefulness of our model.
\begin{table}
\centering
\footnotesize
\caption{Perceived Value of Our model - (Q$_{12}$)}
\label{tab:perceivedvalue}
\begin{tabular}{@{}lcccc@{}}
    \hline
    \textbf{Value} & 0-1 & 2-4 & 5-6 & 7-10 \\
    \hline
    \textbf{\%} & 0 & 8 & 21 & 70 \\
    \hline
\end{tabular}
\end{table}

\begin{table}[h]
    \centering
    \footnotesize
    \caption{Survey Scale for Model Usefulness}
    \label{tab:model_usefulness_scale}
    \begin{tabular}{@{}cp{7cm}@{}}
        \toprule
        \textbf{Score}  & \textbf{Interpretation} \\
        \midrule
        \textbf{0}      & Not useful at all – The model provides no value and does not meet any expectations. \\
        \textbf{1-2}    & Very low usefulness – The model is barely functional or relevant, offering minimal benefit. \\
        \textbf{3-4}    & Low usefulness – Some minor utility, but the model does not sufficiently address user needs. \\
        \textbf{5-6}    & Moderate usefulness – The model is somewhat helpful, but it has limitations that impact its effectiveness. \\
        \textbf{7-8}    & Useful – The model is generally valuable and meets expectations, though some improvements could enhance its impact. \\
        \textbf{9}      & Very useful – The model is highly effective, providing strong value with only minor areas for improvement. \\
        \textbf{10}     & Extremely useful – The model fully meets or exceeds expectations, offering exceptional value. \\
        \bottomrule
    \end{tabular}

\end{table}

\begin{table}[h]
    \centering
    \footnotesize
    \caption{NPS of the Perceived Value of Our model - (Q$_{12}$)}
\label{tab:perceivedvalueNPS}
    \begin{tabular}{@{}lc@{}}
        \toprule
        \textbf{NPS Category} & \textbf{Percentage (\%)} \\
        \midrule
        Promoters (9-10)    & 21.74 \\
        Passives (7-8)      & 47.83 \\
        Detractors (0-6)    & 30.43 \\
        \midrule
        \textbf{NPS} & 39.13 \\
        Interpretation     & Neutral to Good \\
        \bottomrule
    \end{tabular}
    \caption{Distribution of responses according to the Net Promoter Score (NPS) scale}
    \label{tab:nps_distribution}
\end{table}

\begin{table}
    \centering
    \footnotesize
        \caption{Motivation for the chosen value (Table \ref{tab:perceivedvalue}) - (Q$_{13}$)}
    \label{tab:justifications_q12}
    \begin{tabular}{@{}p{7.35cm} r@{}}
        \hline
        \textbf{Motivation} & \textbf{\%} \\
        \hline
        A precise tool would improve effectiveness of addressing \& tracking TD  & 17.39 \\
        Forecasting is useful for proactive management \& prevention & 13.04 \\
        Useful for estimating effort, scheduling \& reducing risks & 4.35 \\
        Mixed opinion (useful but with limitations) & 4.35 \\
        Not useful or low relevance & 0 \\
        \hline
    \end{tabular}

\end{table}

Table \ref{tab:forecasting_preferences_weeks} presents the trade-off between the desired forecast horizon and MAPE, and the percentage of practitioners who preferred each forecast horizon. The findings show a clear trade-off between prediction accuracy and the desired forecasting horizon, capturing the balance that practitioners seek between accuracy and practicability.

The most favored forecast horizons are 4 weeks (1.83\% MAPE, chosen by 48\% of the respondents), 2 weeks (1.44\% MAPE, 30\%), and 6 weeks (2.14\% MAPE, 30\%). They are the most preferred group, where respondents realize the best balance between forecast accuracy and ease of use. Interestingly, the four-week period stands as the most favored duration overall, presumably since it has the smallest prediction error coupled with the length that most closely resembles common sprint lengths for agile methodologies.

Beyond this range, as the forecast horizon extends beyond 8 weeks, MAPE increases linearly, exceeding 3\% at 10 weeks and 5\% at 14 weeks. The declining interest of practitioners with longer forecast horizons indicates that, although longer-term forecasts continue to be provided, the growing uncertainty reduces their functional attractiveness.

Table \ref{tab:forecasting_thematic} classifies the explanations for interval preference forecasting into four general themes. The most common category, Short-Term Preference (33\%), explains that the majority of practitioners prefer shorter forecast horizons (e.g., 2-6 weeks), which is aligned with agile processes and regular technical debt fixing. Suggest an industrial context that requires timely, actionable information instead of long-term forecasts.
The second highest theme, Medium-Term Balance (25\%), implies that there are practitioners who prefer a moderate forecasting period (e.g., 8-14 weeks) and therefore achieve a moderate balance between accuracy and actionable planning. These respondents probably aim to align technical debt planning with milestone-based development cycles.
The accuracy and forecast utility (21\%) contains reasons centered on accuracy and error minimization. These professionals regard precise forecasts with minimal deviation so that decisions made with the model remain actionable and trustworthy.

Technical and Business Constraints (21\%) express external constraints such as hardware obsolescence, product stability, or the feasibility of resolving types of technical debt. This would imply that forecasting needs to be domain-specific and reflect the constraints on the software being developed.

\begin{table}[tb]
    \centering
\footnotesize
    \caption{Practitioners' Preferred Forecasting Intervals and Prediction Accuracy (Weeks) - (Q$_{14}$)}
    \label{tab:forecasting_preferences_weeks}
    \begin{tabular}{@{}rrr@{}}
        \hline
        \textbf{Weeks} & \textbf{MAPE (\%)} & \textbf{\%} \\
        \hline
        2  & 1.44 & 30 \\
        4  & 1.83 & 48 \\
        6  & 2.14 & 30 \\
        8  & 2.96 & 26 \\
        10 & 3.51 & 13 \\
        12 & 4.41 & 4 \\
        14 & 5.09 & 9 \\
        16 & 5.62 & 9 \\
        18 & 6.09 & 4 \\
        22 & 3.00 & 4 \\
        24 & 3.15 & 4 \\
        26 & 3.29 & 4 \\
        28 & 3.43 & 4 \\
        30 & 3.54 & 4 \\
        32 & 3.69 & 4 \\
        34 & 3.84 & 4 \\
        36 & 4.06 & 4 \\
        38 & 4.01 & 4 \\
        40 & 3.84 & 4 \\
        42 & 4.03 & 4 \\
        44 & 4.21 & 4 \\
        46 & 4.39 & 4 \\
        48 & 4.56 & 4 \\
        \hline
    \end{tabular}
\end{table}

\begin{table}[htb]
    \centering
    \footnotesize
    \caption{Thematic Coding of Forecasting Preferences - (Q$_{15}$)}
    \label{tab:forecasting_thematic}
    \begin{tabular}{@{}p{5cm} r@{}}
        \hline
        \textbf{Theme} & \textbf{Percentage (\%)} \\
        \hline
        Short-Term Preference & 33  \\
        Medium-Term Balance & 25  \\
        Accuracy and Forecasting Utility & 21  \\
        Technical and Business Constraints & 21  \\
        \hline
    \end{tabular}
\end{table}

Table \ref{tab:forecasting_months} shows the correlation between the forecast horizons in months, the accuracy of the forecast (MAPE), and the desires of the practitioner. The findings reveal that very accurate near-term forecasts are desired, with 52\% of the practitioners requesting a 1-month forecast (0.30\% MAPE) and 48\% wanting a 2-month horizon (0.34\% MAPE). This reveals that practitioners desire very accurate near-term forecasts for their technical debt planning.

With a growing forecast horizon, preferences continue to decrease, with only 39\% opting for a 3-month forecast (0.50\% MAPE) and 22\% for 4 months (0.61\% MAPE). For more than 6 months, the preferences drop below 13\%, which shows that longer-term forecasts are unrealistic because the error levels rise. The interest is maintained only by the practitioners 9\% for up to 12 months, where MAPE is up to 2.33

Table \ref{tab:forecasting_window_thematic} summarizes the drivers of the selection of the forecast window into four primary themes. The most common theme, Short-Term Preference (30\%), is that practitioners like shorter forecasting windows (1-3 months) so that they can maintain pace with agility and responsiveness in their processes. This is also aligned with sprint-based planning and periodic technical debt pay-down.
Medium-Term Planning (26\%) is the second most frequent theme, and respondents indicated the 1-6 month timescale as the optimal balance of predictability and feasibility in plans. Practitioners generally accept that taking it beyond 6 months makes the forecast less useful because uncertainty increases.
Accuracy and Practicality in Forecasting (22\%) includes those that emphasize the importance of precision in prediction. Practitioners prefer shorter horizons with low MAPE so that the predictions remain valid and practical. They are also concerned that the long-term forecasts are not concrete and practical.
Lastly, Business and Workflow Constraints (22\%) are domain-specific issues such as ensuring product availability, prioritization of backlogs, and customer needs. The development cycles of most practitioners require technical debt to be tackled before large releases or in formalized reporting cycles, most practitioners claim.

\begin{table}[htb]
    \centering
     \footnotesize
    \caption{Practitioners' Preferred Forecasting Intervals and Prediction Accuracy (Months) - (Q$_{16}$)}
    \label{tab:forecasting_months}
    \begin{tabular}{@{}rrr@{}}
        \hline
        \textbf{Months} & \textbf{MAPE (\%)} & \textbf{\%} \\
        \hline
        1  & 0.30 & 52 \\
        2  & 0.34 & 48 \\
        3  & 0.50 & 39 \\
        4  & 0.61 & 22 \\
        5  & 0.73 & 13 \\
        6  & 1.08 & 26 \\
        7  & 1.38 & 13 \\
        8  & 1.61 & 9 \\
        9  & 1.79 & 13 \\
        10 & 1.98 & 9 \\
        11 & 2.15 & 9 \\
        12 & 2.33 & 9 \\
        \hline
    \end{tabular}
\end{table}

\begin{table}[htb]
    \centering
    \footnotesize
    \caption{Thematic Coding of Forecasting Window Preferences - (Q$_{17}$)}
    \label{tab:forecasting_window_thematic}
    \begin{tabular}{@{}p{5cm} r@{}}
        \hline
        \textbf{Theme} & \textbf{\%} \\
        \hline
        Short-Term Preference & 30  \\
        Medium-Term Planning & 26  \\
        Forecasting Accuracy \& Practicality & 22  \\
        Business and Workflow Constraints & 22  \\
        \hline
    \end{tabular}
\end{table}

Table \ref{tab:preferred_td_forecasting} illustrates the intervals preferred by practitioners for forecasting technical debt. The responses are distributed equally in three large bins: Weekly (32\%), Biweekly (32\%) and Monthly (32\%). It mirrors the fact that practitioners prefer frequent forecasting, possibly due to the changing nature of software development and the need to constantly re-estimate technical debt.
Only a minority of 5\% prefer a Sprint-Based Approach (Agile), which suggests that for others, the best forecast window is precisely in synchronization with sprint cycles and not hard-coded time frames. This supports agile practices in which technical debt management is integrated into iterative development phases.

Table \ref{tab:forecasting_window_motivation} categorizes the motivations for the preferred forecasting windows of technical debt professionals. The most common theme, Short-Term Preference (32\%), suggests that most practitioners prefer TD predictions to occur often, which emphasizes the importance of quick feedback loops and the ability to act on evolving issues before they accumulate.
The second most prevalent category, Regular Review and Planning Cycles (27\%), focuses on alignment with existing workflows, such as sprints and formalized reviews. The majority of respondents favor forecast intervals that are aligned with agile iterations so that technical debt can be monitored and dealt with regularly within development cycles.

Balanced Practicality and Granularity (23\%) recognizes drivers who seek a compromise between high frequency and low overhead. Monthly forecasting, for instance, is interruptive enough to be actionable but not so interruptive as to cause problems. Practitioners know that lower frequencies are more precise but are not always practical for a long-term strategy.
Lastly, Domain-Specific Constraints (18\%) incorporate idiosyncratic considerations in terms of practitioners' work environments. This includes energy-efficient systems that must be updated regularly, varied levels of interaction with TD prediction, and other context constraints that affect how frequently TD forecasting is possible.

\begin{table}
    \centering
    \footnotesize
    \caption{Preferred Forecasting Window for TD Prediction - (Q$_{18}$)}
    \label{tab:preferred_td_forecasting}
    \begin{tabular}{@{}p{5cm} r@{}}
        \hline
        \textbf{Preferred Forecasting Window} & \textbf{\%} \\
        \hline
        Weekly & 32  \\
        Bi-Weekly & 32  \\
        Monthly & 32  \\
        Sprint-Based (Agile) & 5  \\
        \hline
    \end{tabular}
\end{table}

\begin{table}
    \centering
    \footnotesize
    \caption{Thematic Coding of Forecasting Window Motivations - (Q$_{19}$)}
    \label{tab:forecasting_window_motivation}
    \begin{tabular}{@{}p{5cm} r@{}}
        \hline
        \textbf{Theme} & \textbf{\%} \\
        \hline
        Short-Term Preference & 32  \\
        Regular Planning and Review Cycles & 27  \\
        Balanced Practicality \& Granularity & 23  \\
        Domain-Specific Constraints & 18  \\
        \hline
    \end{tabular}
\end{table}

\end{ReviewerAEnv}

\section{Discussion}
\label{sec:Discussion}

The objective of the presented study was to analyze the impact of temporal factors in code TD prediction. Code TD was approximated using the SQALE index metric calculated by SQ, and the predictions were made including the set of different types of code smell issues in the projects within the prediction models. The adoption of TSA prediction models enabled empirical demonstration of the suitability of time-dependent techniques for code TD prediction, as well as comparative analysis with commonly used ML prediction algorithms. Consequently, 11 different prediction models were constructed for each project. The analysis included a comprehensive data collection, preprocessing, and examination of 31 open-source Java projects. However, to ensure the integrity and fairness of the results, only the outcomes of the 14 projects that provided complete results with the adopted prediction models were considered.  


\begin{ReviewerAEnv}
Our study aimed to answer six research questions. The first research question aimed to compare the predictive performance of two time-dependent approaches based on ARIMA already implemented in the academic literature given the same experimental setting. The confronted models were the ARIMAX model previously implemented by Mathioudaki et al.~\cite{Mathioudaki2022}, and the ARIMA+LM model previously implemented by Zozas et al.~\cite{Zozas2023}. Within the same setting for our first research question, we also assumed before running the study that we expected the accuracy from the ARIMAX model, given its existing theoretical background and the factor that it considers as a model variable the autoregression of its dependent variable, would be superior to that of the ARIMA+LM, which comprises the mixture of two different families of statistical models, and performs a linear regression on the dependent variable without considering its autoregressive component as an additional variable within the model. Thus, looking at the results, our initial takeaway depicted the superiority of ARIMAX over the ARIMA+LM model in predicting the code TD. These results suggest that, given a time-dependent prediction context, \textbf{ multivariate models that include and simultaneously predict independent variables as well within the prediction model, such as ARIMAX, succeed on providing robust results} as shown in Figure~\ref{fig:rq1_results}.
\end{ReviewerAEnv} 

Furthermore, in the second research question, we proved the robustness reported with ARIMAX by comparing its Code TD prediction performance against a set of the most common ML algorithms adopted in the SE field. The results provided in Tables~\ref{tab:monthlyPerformance} and~\ref{tab:biweeklyPerformance} corroborated this evidence. These results suggest the necessity of adopting models that consider the time dependence factor in the dependent variable to be predicted. However, it is worth noticing the close results reported by models such as Random Forest or Extreme Gradient Boost, prediction algorithms widely used in academia as well as in industry. Given the non-linear distributional assumption of these models and the oppositely linear distributional assumption, which is the foundation for TSA models such as ARIMAX, \textbf{leveraging the combination of these models through ensemble learning approaches might further improve the prediction performance reported in this study}.

In addition, we analyzed the impact of addressing the seasonal component within the time-series data on the obtained results. For that, we used the model extensions provided in ARIMA-founded models to analyze the results for the SARIMAX and SARIM + LM models. As a result of the analysis performed, we observed slight improvements in the initially adopted TSA models, thus depicting \textbf{a small impact on prediction performance by controlling the seasonality component}. This evidence suggests a small dependence on seasonality from the perspective of the OS software development community, which might not follow structured development schedules compared to the software development activity performed in the industry. 

As the key takeaway of this study,  we can confirm that time-dependent models \textbf{ are applicable and competitive in predicting the TD code}, not only at the level of a single project, but also at the aggregate level, as we have demonstrated. More research is still needed to understand the methodology required to prepare the most suitable data for TSA models. It has been demonstrated that if the data fits well in the TSA model, the latter can provide a high prediction performance. Therefore, this study suggests \textbf{the adoption of periodic software quality analysis practices within the project development process, thus providing a complete monitoring of metrics such as Code TD and better quality data for the improvement of prediction performance}. Similarly, it is important to note at this stage that when we make our predictions, these values are the obtained most likely estimates; therefore, we are not comparing our real values against the result of a simulation or an execution.  
TSA models can have a considerable prediction performance, but there is a need for further research in modeling Code TD prediction while considering time as a relevant factor.

\subsection{\ReviewerA{Forecasting Code TD}}
\begin{ReviewerAEnv}
The observed results from the long-term Code TD forecasting performance resulted in a set of key takeaways. Through the analysis implemented that addresses long-term forecasting, we could confirm \textbf{that time-dependent models can provide competitive code TD forecasting for single-step and multistep forecasting scenarios}, and therefore highlight \textbf{the importance of considering techniques that treat the temporal nature of the dependent variable} when performing data analysis on time-dependent variables. Still, more work needs to be done on long-term code TD forecasting to provide robust enough long-term maintenance guarantees to practitioners, especially given the volatility advertised in the variance observations of the results we obtained.
Similarly, practitioners should benefit from different approaches of Code TD prediction, i.e. single-step forecasting or multi-step forecasting, based on the specific characteristics and needs of the software project workflows. On this basis, our work emphasizes \textbf{the benefits of structuring Code TD analysis executions in a serialized style} when working on maintenance of the software quality with tools like SQ. 
\end{ReviewerAEnv}

\begin{ReviewerAEnv}
    
\subsection{Impact on Practitioners}
We surveyed the opinion of \textbf{23} practitioners and their expertise profile. We asked them to evaluate the practical usability of our model and the perceived benefit or limitations it may have in an industrial context. Referring to our findings in the RQ$_5$, we presented the practitioner with the same time window of 36 months and 72 weeks. In particular, no one exceeded the 48-week mark, while only 4\% on average selected a time window beyond 12 weeks. Furthermore, no instances surpassed the 12-month threshold.

According to Tables \ref{tab:perceivedvalue} and Table \ref{tab:justifications_q12} describe the external validity of our TD prediction model, with overwhelming percentages of respondents knowing its ability to improve TD management. The high ratio of Promoters and Neutral-to-Promoter respondents indicates our approach is well perceived, justifying its use in real software engineering practice.
Table \ref{tab:forecasting_preferences_weeks} confirms the precision of our TD forecasting model within short- to medium-term horizons, specifically the 2-6 week horizon, where precision is best and trust from practitioners highest.

Generally speaking, our survey findings confirm that our forecasting model best suits industry needs, particularly in the case of short- to medium-term planning, where practitioners tend to prefer its forecasts. In fact, according to Table \ref{tab:forecasting_thematic}, practitioners favor a good accuracy with a medium time window over the planning horizon, where the ideal is between 1 and 3 months. This aligns perfectly with typical agile development cycles and decision-making milestones, confirming that our approach is meaningful within substantial periods in industry.

Moreover, Table \ref{tab:forecasting_window_thematic} validates the accuracy of our model's performance, particularly in short- to medium-term forecasting, which has been widely accepted to be informative, accurate, and actionable for real software development processes. Table \ref{tab:preferred_td_forecasting} reiterates that short-range forecasting is crucial to practitioners because it places technical debt in continuous observation and is resolved within appropriate actionable time frames. The lack of preference for longer-range forecasting (e.g., quarterly or yearly) also emphasizes the need for near-term and actionable forecasts over long-horizon predictions.

Finally, Table \ref{tab:forecasting_window_motivation} shows that our forecasting approach aligns more closely with what the industry desires, particularly in short- to medium-term window lengths, where usability, precision and integration are the most valued by practitioners.

\end{ReviewerAEnv}

\section{Threats to Validity}
\label{sec:Threats}

\paragraph{Construct validity} Some projects have been affected by only a few code smell rule issues, so we decided to exclude them because their independent variables were considered uninformative. 
Additionally, irregularities in commit data can cause missing data for the generated periodical time series. The threat was minimized by linearly interpolating the values for the missing periods. Approximations cannot be as accurate as using real data; therefore, we could expect different results if periodic complete real data existed.
\ReviewerA{An additional potential threat to construct validity is identification and TD quantification, namely to the validity of code smells as predictors for TD. Although SonarQube's remediation time is widely used, other work \cite{esposito_can_2023,Lenarduzzi2019,esposito_extensive_2024} demonstrated that it is not always a good estimator of the actual effort to remove TD. So, our results may be affected by potential TD measurement errors. To restrict this risk, we compared our approach with existing research on TD assessment and ensured that our approach was consistent with current best practice. However, we acknowledge that variability in the estimation of TD is a feature of automated analysis and that further research should explore other methods of quantification of TD to further validate our results.}

\paragraph{Internal validity} 
\ReviewerA{We selected the SQALE index as the variable of interest, given its capacity to depict technical debt. We then established a set of code smell violation rules reported by SQ as the underlying reason for the existence of the given SQALE index metric. This served as the initial set of independent variables in our study. However, we did not consider collecting additional variables that could have an impact on the results. Consequently, alternative choices of independent variables may yield different results. Concerning the impact on the remediation time estimate provided by the collected SQALE index data, previous research has demonstrated that the SQALE index provides overestimations of the real code TD. Therefore, this study uses the SQALE index as a proxy for technical debt, as well as the independent variables for code smell rule violations in our models. We acknowledge that potential overestimations in the SQ estimated SQALE index data could affect the practical applicability of the study findings. In future research, we will mitigate this threat by incorporating more accurate estimates of remediation time into models, or complementing the models with real-effort data to improve the model's reliability.}

\paragraph{External validity} The study subjects were mature open source projects written in Java that met the rigorous quality standards of ASF. 
The included projects represent a wide spectrum of application domains, including external libraries, frameworks, web utilities, and substantial computational infrastructures. Hence, the obtained results can be generalized to projects with similar characteristics, and therefore nonmature projects, retired projects, and projects not using SonarQube are excluded. 

\paragraph{Conclusion validity} This study applied commonly known statistical and ML techniques. During data analysis, we ensured that the assumptions of the techniques were fulfilled. However, a low number of data points in some projects can reduce the prediction power of the models used. Additionally, the ARIMA + LM model may not capture the time-dependent nature of the SQALE index, as its prediction is performed through a Multiple Linear Regression model, which does not consider the time-dependent factor.
Regarding the reliability of the measures, the variables' measurements were collected from the data set and the construction of the final set of model variables was carried out through an automated process. For the sake of the reliability of the results, we excluded the collected projects which did not allow the algorithms from the adopted models to converge within the model fitting. Therefore, we report only the aggregated results from the projects that provided results for the entire set of models considered in this study.

\section{Related Work}
\label{sec:RelatedWork}
\begin{table*}[htb!]
\centering
\caption{Related work on the existing research on TD prediction}
\label{tab:relatedworks}
\resizebox{\linewidth}{!}{%
\begin{tabular}{m{3.8cm} | m{3.5cm}| m{3.5cm} |m{3.5cm} |m{3.5cm}| m{3.5cm} |m{3.5cm} | m{3.5cm} }
\hline 
 & \textbf{Tsoukalas et al. \cite{Tsoukalas2020}} & \textbf{Tsoukalas et al. \cite{tsoukalas2021machine}} & \textbf{Mathioudaki et al. \cite{Mathioudaki2021}} & \textbf{Aversano et al. \cite{Aversano2022}} & \textbf{Mathioudaki et al. \cite{Mathioudaki2022}} & \textbf{Zozas et al. \cite{Zozas2023}} & \textbf{Our work}  \\\hline 
\textbf{Seasonality} &  &  &  &  &  &  & \ding{51} \\\hline 
\textbf{Time Series (TS) processing methodology} & Weekly snapshots across 3 years \& Walk-Forward Train-Test & No serialization \& 10-K CV (No TS methodology) & Sliding Window over project commits \& Walk-Forward Train-Test & Raw commit-level observations \& Repeated CV (No TS methodolofy) & Last commit of the week as weekly observation \& Walk-Forward Train-Test & Raw release data \& Random train-test split \& Walk-Forward Train-Test & Biweekly and Monthly serialization of SQ analyses \& Walk-Forward Train-Test\\ \hline 
\textbf{Feature selection} & Correlation analysis & Hypothesis testing &  & Correlation analysis &  & BSR & Feature importance, Variance thresholding, Zero percentage, Correlation analysis \\ \hline 
\textbf{Hyperparameter tuning} & Grid-search method & Grid-search method & Grid-search method &  & ACF, PACF & ACF, PACF & Auto-ARIMA \\\hline 

\textbf{Univariate models} &  & Univariate logistic regression & Multi-layer perceptron &  & ARIMA & ARIMA & ARIMA, SARIMA \\\hline 

\textbf{Multivariate models} & MLR, L1, L2, SGD, SVR (linear), SVR (rbf), RF (regression) & LR, NB, DT, KNN, SVM, RF, XGB &  & MLR, BDT, RF & ARIMAX & LM & ARIMAX, SARIMAX, ARIMA+LM, SARIMA+LM L1, L2, MLR, SGD, SVR, RF, XGB \\\hline 

\textbf{Industrial survey} & \ding{51} &  &  &  &  & \ding{51} & \ding{51} \\\hline 

\textbf{Long-term forecasting} & 1 to 40 weeks &  & 5 to 150 commits &  & 4, 8, 12 steps & 12 iterations & 1 to 36 months \\\hline 

\textbf{Time Series models} &  &  & ARIMA &  & ARIMA, ARIMAX & ARIMA & ARIMAX, ARIMA+LM, SARIMAX, SARIMA+LM \\\hline 

\textbf{ML models} & MLR, L1, L2, SGD, SVR (linear), SVR (rbf), RF (regression) & LR, NB, DT, KNN, SVM, RF, XGB & RF & MLR, BDT, RF &  &  & L1, L2, MLR, SGD, SVR, RF, XGB \\\hline 

\textbf{DL models} &  &  & Multi-layer perceptron &  &  &  &  \\\hline 

\textbf{Cross-validation} & Walk-Forward Train-Test & Repeated stratified CV & Walk-Forward Train-Test & Repeated CV & Walk-Forward Train-Test & Walk-Forward Train-Test & Walk-Forward Train-Test \\\hline 

\textbf{Performance metrics} & MAPE & Precision, Recall, F2, AUC, Module inspection & MAPE, MAE, RMSE & RMSE & MAPE, MAE, RMSE & MAPE, MAE, RMSE & MAPE, MAE, RMSE \\\hline 

\textbf{Sample size} & 15 projects & 25 projects & 5 projects & 8 projects & 5 projects & 105 projects & 14 projects \\\hline 

\textbf{Results discussed} & \ding{51} & \ding{51} & \ding{51} & \ding{51} & \ding{51} & \ding{51} & \ding{51} \\ \hline 
\textbf{Scripts available} & \ding{51} & \ding{51} &  &  &  &  & \ding{51} \\\hline 
\textbf{Datasets available} & \ding{51} & \ding{51} &  & \ding{51} &  & \ding{51} & \ding{51} \\
\hline 
\end{tabular}%
}
\end{table*}

\begin{ReviewerBEnv}
This section reports related work, including the current state of Technical Debt (code TD) prediction and a detailed comparison with our work in Table~\ref{tab:relatedworks}.

Tsoukalas et al.~\cite{Tsoukalas2020} investigated the effectiveness of ML techniques in modeling and predicting the evolution of the code TD. Specifically, they collected weekly commit-level snapshot observations over three years for a total of 15 software projects. They used the correlation analysis technique to perform feature selection and the grid search method to tune the model hyperparameters. They implemented a set of different well-known ML models (see Table~\ref{tab:relatedworks}) and trained and tested them through the Walk-Forward (WF) Train-Test method, considering MAPE as a performance assessment metric. They provided the results, scripts, and data sets for the study. Similarly, they introduced an industrial survey to empirically assess the significance of the TD approach introduced in their work.

In a second work, Tsoukalas et al.~\cite{tsoukalas2021machine} adopted a different set of ML models to classify code TD classes (High/Not-High TD), analyzing 25 open-source projects. They performed hypothesis testing for the selection of the most significant independent variables and tuned the parameters of the selected models through the grid-search method. In this second work, they used commonly known classification assessment methods such as AUC, F2 or Recall, among others, in a repeated stratified cross-validation setting to assess their results. Similarly, as in their previous work, they shared the study's results, scripts, and datasets.

Mathioudaki et al.~\cite{Mathioudaki2021} explored the application of deep learning methods (DL) compared to univariate approaches already implemented such as ARIMA or RF models to improve the precision of long-term code prediction. TD. To achieve this, they created, evaluated, and juxtaposed DL models using a data set that includes five prominent real-world software applications sourced from the Technical Debt dataset version 1~\cite{LenarduzziPromise2019}. They preprocessed the adopted commit-level data through the Sliding Windows technique to transform the raw data into a processable time series for the chosen models. They also adopted the Grid Search method for model parameter tuning. The long-term prediction horizon ranged from 5 to 150 commit-ahead lengths. To assess the performance of the predictions, they considered the WF Train-Test technique and the MAPE, MAE, and RMSE performance metrics. The results revealed that DL techniques produce code TD prediction models with commendable predictive accuracy, extending up to 150 steps ahead in future prediction.

Aversano et al.~\cite{Aversano2022} investigated the potential utility of software system quality metrics to accurately predict the TD code. The authors examined quality metrics from 8 distinct open source software systems and then fed those commit-level metrics into different ML algorithms to predict TD. They performed a correlation analysis for the selection of the features of the independent variables used in their models. They used the repeated cross-validation technique to assess their results, which were quantified through the RMSE. The results demonstrated strong predictive performance, and the suggested approach offers a valuable method to comprehend the practical aspects of the technical debt phenomenon.

Focusing on the work that adopted TSA models to predict code TD, we concentrated on two papers. 
Mathioudaki et al.~\cite{Mathioudaki2022} investigated the predictive capabilities of TSA models for the prediction of the code TD. They aimed to determine whether the incorporation of independent variables as predictors, known as Code TD predictors, into ARIMAX models~\cite[p.~451]{chan2008time} could produce more precise Code TD predictions compared to conventional univariate Autoregressive Integrated Moving Average (ARIMA) models. Their investigation used five datasets that captured the historical evolution of software metrics derived from static code analysis in five long-standing projects. The raw data was serialized into weekly observation points by choosing the last commit performed in each weekly time window. They used Auto-Correlation Functions (ACF) and Partial Auto-Correlation Functions (PACF) for the parameter tuning of the chosen models. They generated predictions using both ARIMA and ARIMAX models for various time horizons on these data, exploring long-term forecasting horizons of 4, 8 and 12-week ahead periods. The results yielded a clear conclusion. Across the open-source software projects examined, the accuracy of the ARIMAX models exceeded that of the ARIMA models by a significant margin.

More recently, Zozas et al.~\cite{Zozas2023} combined two models to predict code TD: Supervised Linear Regression (LM) and ARIMA. The authors used backward stepwise regression (BSR) to identify the most significant predictors to describe the TD response variable code. They used Auto-Correlation Functions (ACF) and Partial Auto-Correlation Functions (PACF) for the parameter tuning of the chosen models. They performed a univariate TSA prediction through ARIMA for each of the selected predictors to predict future values. Then, they used the regression model previously developed to estimate the future values of the response variable. To assess the performance of their models, they adopted the MAPE, MAE, and RMSE performance metrics within a WF Train-Test setting. They made 12-week long-term predictions ahead to explore the long-term prediction performance of the models studied. The results looked promising as an additional approach to predict TD efficiently, and they provided the results and data sets from their work. In addition, they organized a short survey with 15 JavaScript developers with medium experience to evaluate their findings.

In our work, instead of comparing univariate and multivariate TSA, we aim to directly compare different multivariate TSA approaches to investigate which approach obtains better prediction performance. Furthermore, we explore the impact of addressing seasonality as a critical component impacting the performance of TSA models on code TD prediction, and confront the performance of the considered TSA multivariate models towards a set of previously implemented ML techniques used for code TD. The selected predictors will be used to predict the value of the SQALE index, which will be our target value. Our approach relies on performing different multivariate applications of the ARIMA model. To test its performance, we aim to split the data for training and testing the model through the WF Train-Test approach and performance metrics such as MAPE, MAE and RMSE. We also aim to perform long-term forecasting with the best resulting TSA models with forecasting windows reaching the 3-year horizon. Moreover, we are interested in what preferred code TD forecasting horizon length, and we collected opinions from practitioners in a shared industrial survey. 

Furthermore, since training ML models with a single feature can compromise their quality \cite{esposito2023uncovering}, we determine the optimal number of features and compare the performance of selected multivariate TSA models against a set of ML models chosen based on previous research.
\end{ReviewerBEnv}

\section{Conclusions}
\label{sec:Conclusion}

We conducted an empirical study to compare two TSA approaches, their respective version adjusted to treat the seasonality component, and seven ML algorithms to code the prediction of TD as the SQALE index calculated by SQ. In addition, we conducted an industrial survey to empirically analyze the implications of our findings from the perspective of practitioners and tested the accuracy of the resulting best time-dependent models when performing long-term forecasting.

We trained the compared methods using Code TD observation data that were serialized into biweekly and monthly periodicity levels from the resulting number of 14 open source Java projects. The comparison denoted a clear superiority of the ARIMAX model compared to the other adopted models when training the models with biweekly time series. At the same time, SARIMAX provided better results with monthly time series. However, the results provided by SARIMAX suggested that the impact of seasonality could not substantially improve the predictive performance of ARIMAX, thus demonstrating that the ARIMAX model generates the main improvement between the performance of the ML algorithms and the TSA models. Our findings show that seasonality has little impact on the predictive performance of time-dependent techniques. However, incorporating time dependence improves the predictions over time-agnostic methods. Moreover, coupled with our data serialization process, it highlights the benefits of sequential software quality analysis in improving Code TD prediction.

Using the TSA models with the resulting best predictive performance, we performed long-term forecasting up to a forecasting horizon of three years translated into 36 months and 72 biweekly periods accordingly. The resulting forecasts looked promisingly balanced, which reaffirms the suitability of time-dependent techniques for Code TD predictions. Finally, our survey reveals good industry confidence in our tool and its effectiveness for short- to medium-term forecasting windows, which aligns closely with industry needs and agile methodology practice.

In future work, our aim is to analyze the implementation of transformer-driven prediction techniques, which allow the inclusion of low-level models into the multiple deep learning layers nested within the transformer architecture. Adapting TSA techniques with these models provides great modeling ability for long-range dependencies and interactions in sequential data and TSA models, thus further contributing to the research community with advanced predicting techniques and anticipating Code TD.

\section{Data Availability Statement}
\label{sec:Replicability}

\begin{ReviewerAEnv}
To allow verifiability and replicability, we made the raw data and analysis files available in our online appendix. Furthermore, we provide instructions on how to use the released replication package in~\ref{appendix:replication}, as well as the README file in our online appendix.\footnote{\label{package} https://doi.org/10.5281/zenodo.14974421}
\end{ReviewerAEnv}

\bibliographystyle{model1-num-names}
\bibliography{main.bib}
\appendix

\section*{Appendix}
In this Section, we detail parameter tuning of the Box-Jenkins models and how to replicate our study.

\section{Parameter tuning of the Box-Jenkins models}
In this section of the appendix we describe the mathematical definitions of the model parameters composing the TSA models included in the work, \ReviewerA{as well as the summary of the steps to be followed in order to implement model parameter tuning of the \textit{Box-Jenkins} models. The models adopted in this study are multivariate versions of the classic Box-Jenkins model}, better known as ARIMA~\cite{box2015time}, \ReviewerA{therefore their theoretical background reside on the definition of the ARIMA model}. This model is a univariate TSA model composed of the autoregressive component, the moving average component, and the differencing component, as explained below.

\subsection{Autoregression (AR) order (\textit{p / P} parameter)}

The autoregressive process of order \textit{p}, AR(\textit{p}), denotes a stationary time series $\left\{X_t\right\}_{t \in Z}$ such that 
\begin{equation}
    X_t = \sum^p_{i=1}\phi_iX_{t-i} + \varepsilon_t
\end{equation}
where $\left\{\phi_1, \cdots, \phi_p\right\}$ are fixed autoregression coefficients and $\left\{\epsilon_t\right\}$ are independent random noises of mean $\mu_0$ and constant variance $\sigma^2$. An AR(\textit{p}) process is \emph{stationary} if the \textit{p} roots of $\phi(z) = 1 - \phi_1z - \phi_2z^2 - \cdots - \phi_pz^p$ fall within the random state unit circle~\cite{box2015time}. 

Seasonality adjustment occurs when the autoregressive pattern occurs at multiple lags of the defined periodicity (\textit{m}). Thus, a seasonal autoregression (SAR) process is defined by 
\begin{equation}
    X_t = \sum^p_{i=1}\Phi_iX_{t-S} + \varepsilon_t
\end{equation}
where $\Phi$ is defined as $\Phi(z) = 1 - \Phi_1z - \Phi_2z^2 - \cdots - \Phi_Pz^P$.

\subsection{Moving Average (MA) order (\textit{q / Q} parameter)}

The moving average process of order \textit{q}, MA(\textit{q}) denotes a stationary time series $\left\{X_t\right\}_{t \in Z}$ such that
\begin{equation}
    X_t = \sum^q_{k=1}\theta_k\varepsilon_{t-k}
\end{equation}
where $\epsilon_t \sim N(0, \sigma^2)$ and $\theta$ are defined as $\theta(z) = 1 + \theta_1z + \theta_2z^2 + \cdots + \theta_qz^q$. Every MA(\textit{q}) process is stationary, but there can be two different MA(\textit{q}) processes with the same autocovariance~\cite{brockwell2002introduction}. We avoid this issue by enforcing invertible MA processes (see details in Section~\ref{sec:Replicability}). 

As before, a seasonality adjustment is performed when the moving average pattern occurs at multiple lags of the defined periodicity (\textit{m}). Thus, seasonal moving average (SMA) processes are defined by 
\begin{equation}
    X_t = \sum^q_{k=1}\Theta_k\varepsilon_{t-k}
\end{equation}
where $\Theta$ is defined as $\Theta(z) = 1 + \Theta_1z + \Theta_2z^2 + \cdots + \Theta_Qz^Q$.

\subsection{Differecing order (\textit{d / D} parameter)}

Among the existing approaches to reach stationarity, models such as ARIMA apply differencing repeatedly to the series of data until the differenced observations resemble a realization of stationary time series, required for accurate forecasting~\cite{brockwell2002introduction}. Considering, for instance, an original time series $\left\{X_t\right\}$ we can define the first differencing order (\textit{d} = 1) as:
\begin{equation}
    Y_t = \nabla X_t = X_t - X_{t-1}
\end{equation}

Hence, the differencing operator or lag-\textit{d} can be formulated as
\begin{equation}
    \nabla_d X_t = X_t - X_{t-d}.
\end{equation}
Moreover, differencing can also be used in cases where the seasonality of the series is addressed. If the series has a seasonal pattern, then \textit{d}-lag differencing order can remove this pattern. Thus, if \(X_t = T_t + S_t + \varepsilon_t\) and $S_t$ had period \textit{d}, then:
\begin{equation}
    \nabla_d X_t = T_t - T_{t-d} + \nabla_d \varepsilon_t
\end{equation}
where $S_t$ stands as the seasonality component at series observation \textit{t}, $T_t$ is the trend components accordingly, and $\varepsilon_t$ is the existing random noise.

\subsection{Model parameter tuning}
\ReviewerA{
The combination of the model parameter components explained previously in this appendix results in the following formula, which we provide in its multivariate format by including the $X_{t,i}$ independent variables  given the claim of our work supporting this type of models:}

\ReviewerA{
\begin{equation}
    Y_t = \phi_0 + \sum^k_{i=1}\beta_iX_{t,i} + \sum^p_{j=1}\phi_jX_{t-j} + \sum^q_{k=1}\theta_k\varepsilon_{t-k} + \varepsilon_t
\end{equation}}

\ReviewerA{where:}

\ReviewerA{
\begin{itemize}
    \item $\phi_0$ is a model constant.
    \item $\phi_j$ and $\theta_k$ are the AR and MA coefficients. 
    \item $\beta_i$ are the regression coefficients for the $X_{t,i}$ independent variables included after the backward variable selection criteria.
    \item $\varepsilon_t \sim N(0, \sigma^2)$.
\end{itemize}
}

\ReviewerA{This ARIMAX formula can be easily transformed (i) into an univariate ARIMA model, by removing the independent variable regression component and (ii) into seasonally adjusted models such as the studied SARIMAX by integrating the seasonal components ($P,Q,D$) into the regression.
For each combination of independent variables, this study leveraged the \textit{Auto-arima} algorithm as mentioned in Section 3.5.3, to tune the model parameters. This algorithm automates the autoregression and moving average components to provide optimal forecasting results. Initially, it identifies the optimal differencing ($d$) order to provide stationarity to the time series. Its stepwise process identifies the best AR and MA combinations, respectively. For each possible model parameter combination, it assesses its estimation capability by measuring the \textit{Akaike Information Criterion} (AIC)~\cite{fuller2009introduction} and \textit{Bayesian Information Criterion} (BIC)~\cite{fuller2009introduction}, both common measures to assess the relative quality of statistical models.
The AIC assesses the model goodness of fit based on the estimated likelihood, while it includes a penalty on the increase of the number of parameters included in the model to prevent over-fitting. This could be defined as follows:}

\ReviewerA{
\begin{equation}
    AIC = 2k - 2 ln(\hat{L})
\end{equation}}

\ReviewerA{where $k$ is the number of included parameters and $L$ is the estimated likelihood.
}

\ReviewerA{Similarly to the AIC, the BIC criterion penalizes the increase of model parameters to prevent model over-fitting, while also accounting for the sample size as a component in the calculated penalty, thus favoring stronger prevention of over-fitted models:}

\ReviewerA{
\begin{equation}
    BIC = k ln(n) - 2 ln(\hat{L})
\end{equation}}

\ReviewerA{where k is the number of included parameters, n is the sample size and L is the estimated likelihood.}

\ReviewerA{Therefore, the stepwise parameter tuning search is repeated until the model cannot be further improved. Consequently, this process generates a model with the most efficient set of independent variables and model parameters. Table~\ref{tab:parametersummary} provides a summary containaining the resulting best ARIMAX and SARIMAX model parameters ($p, d, q$) and ($P, D, Q, m$), as selected for each project, thus facilitating the reproducibility of our results. We also provide the table with the best model parameter combinations for models ARIMA+LM and SARIMA+LM univariate models in the published replication package.}

\section{\ReviewerB{Replication of the study}}
\label{appendix:replication}

\ReviewerB{In this section of the appendix, we explain the instructions needed to replicate the work of this study. For that, we will emphasize the needed requirements to execute the replication code, the architecture of the shared code as well as the possible code execution options the reader can opt to if wished.
The codes used to run this study were written in Python 3.9. We provide instructions on how to install a virtual environment where to install the necessary dependencies to run the code, these are also provided in a requirements file within the replication package. We used well-known statistics and ML libraries such as \textit{statsmodels}, \textit{scikit-learn} and \textit{scipy} to implement the selected models. Additionally, we used the \textit{pmdarima} library to implement the \textit{Auto-arima} algorithm.} 

\ReviewerB{The code is organized and structured in a format that facilitates replication for practitioners. All the Python modules are connected to the \textit{main} module. Therefore, practitioners only need to run this module to replicate our work. Moreover, if practitioners are interested in key parts of the implementation, the \textit{commons} module allows them to activate and deactivate sections of the program execution so that these are skipped. We provide further instructions on the program execution sections and a detailed implementation description in the online appendix.}

\onecolumn
\scriptsize
\begin{longtable}{llllllllllll} 
\caption{Summary table containing the resulting best ARIMAX/SARIMAX model parameters ($p, d, q$) and ($P, D, Q, m$).} \label{tab:parametersummary} \\

\hline
\textbf{Project} & \textbf{Model} & \textbf{\begin{tabular}[c]{@{}l@{}}Seriali\\ zation\end{tabular}} & \textbf{p} & \textbf{d} & \textbf{q} & \textbf{P} & \textbf{D} & \textbf{Q} & \textbf{m} & \textbf{AIC} & \textbf{Regressors} \\ 
\hline
\endfirsthead

\hline
\textbf{Project} & \textbf{Model} & \textbf{\begin{tabular}[c]{@{}l@{}}Seriali\\ zation\end{tabular}} & \textbf{p} & \textbf{d} & \textbf{q} & \textbf{P} & \textbf{D} & \textbf{Q} & \textbf{m} & \textbf{AIC} & \textbf{Regressors} \\ 
\hline
\endhead

\hline
\multicolumn{12}{r}{\textit{Continued on next page}} \\ 
\hline
\endfoot

\hline
\endlastfoot

archiva          & Arimax         & monthly                                                           & 0          & 0          & 1          &            &            &            & 12         & 777.97       & \begin{tabular}[c]{@{}l@{}}S1213, S1192, RedundantThrowsDeclarationCheck, S00117, \\ S1488, DuplicatedBlocks, S1186, S1132\end{tabular}                                                                                                                     \\
bcel             & Arimax         & monthly                                                           & 1          & 0          & 1          &            &            &            & 12         & 1194.46      & \begin{tabular}[c]{@{}l@{}}S1213, ModifiersOrderCheck, S1192, S00117, S00122, S1488, \\ DuplicatedBlocks, S134, S1066, MethodCyclomaticComplexity, \\ S00112, S00108, S1186, S1151, S1132\end{tabular}                                                      \\
codec            & Arimax         & monthly                                                           & 0          & 1          & 1          &            &            &            & 12         & 1312.76      & \begin{tabular}[c]{@{}l@{}}ModifiersOrderCheck, S1192, RedundantThrowsDeclarationCheck, \\ S00117, DuplicatedBlocks, S134, MethodCyclomaticComplexity, \\ S1226, S00112, S00108, S1186, S1151, S1132\end{tabular}                                           \\
collections      & Arimax         & monthly                                                           & 0          & 0          & 1          &            &            &            & 12         & 1634.78      & \begin{tabular}[c]{@{}l@{}}S1213, ModifiersOrderCheck, S00117, S00122, S1488, \\ DuplicatedBlocks, S134, MethodCyclomaticComplexity, \\ S1226, S00112, S1155, S00108, S1186, S1151, S1132\end{tabular}                                                      \\
\end{longtable}
\twocolumn

\begin{table*}
    \caption{Descriptive statistics from the projects used in the analysis.}
    \label{tab:descriptiveStats}
    \footnotesize
    \centering
    \setlength{\tabcolsep}{4pt}
    \resizebox{0.9\textwidth}{!}{
    \begin{tabular}{@{}p{1.3cm}p{1.25cm}p{0.4cm}lll@{\extracolsep{4pt}}llll@{\extracolsep{4pt}}llll@{}}
        \toprule
        \multirow{2}{*}{\shortstack{\textbf{Apache}\\ \textbf{projects}}} & \multirow{2}{*}{\shortstack{\textbf{Time}\\ \textbf{frame}$^\diamond$}}  & \multicolumn{4}{c}{\textbf{Original data}} & \multicolumn{4}{c}{\textbf{Monthly data}} & \multicolumn{4}{c}{\textbf{Biweekly data}}\\
        \cmidrule{3-6} \cmidrule{7-10} \cmidrule{11-14}
         & & \textbf{\#} & \textbf{Min/Max \textit{SI}} & \textbf{Mean \textit{SI}} & \textbf{Std \textit{SI}} & \textbf{\#} &  \textbf{Min/Max \textit{SI}} & \textbf{Mean \textit{SI}} & \textbf{Std \textit{SI}} & \textbf{\#} & \textbf{Min/Max \textit{SI}} & \textbf{Mean \textit{SI}} & \textbf{Std \textit{SI}}\\
         \midrule
        \textit{archiva} & 05/11-02/12 & 1,630 & 0\textminus172,634 & 71,068 & 34,803 & 65 & 0\textminus116,422 & 68,924 & 32,639 & 136 & 0-121,632 & 68,916 & 31,354\\
        \textit{batik} & 10/00-08/02 & 941 & 65,831\textminus175,362& 132,111 & 29,957 & 21 & 71,725\textminus174,077 & 138,100 & 30,581 & 44 & 71,725\textminus174,968 & 140,551 & 29,246\\
        \textit{bcel} & 10/01-04/18 & 291 & 47,854\textminus59,608 & 54,304 & 3,262 & 173 & 47,869\textminus58,821 & 54,348& 3,223 & 394 & 47,869\textminus58,821 & 54,426 & 3,233\\
        \textit{beanutils} & 03/01-10/18 & 324 & 1,380\textminus58,250 & 39,025 & 14,346& 190 & 1,380\textminus58,142 & 47,057 & 12,111 & 427 & 1,380\textminus58,142 & 47,154 & 11,839\\
        \textit{cayenne} & 01/07-07/08 & 353 & 179,813\textminus223,285 & 205,411 & 12,555 & 17 & 179,813\textminus222,828 & 203,314 & 13,481 & 33 & 179,813\textminus222,875 & 203,402 & 13,344\\
        \textit{cocoon} & 03/03-02/07 & 2,311 & 64,810\textminus286,277 & 183,528 & 65,634 & 46 & 66,640\textminus282,280 & 181,404 & 75,407 & 95 & 65,370\textminus283,313 & 179,951 & 73,980\\
\end{tabular}
    }
\end{table*}

\begin{table*}[] 
    \caption{Resulting and discarded projects from the model training stage. (\textbf{Bold}: Projects included in the final results.)}
    \label{tab:resultingprojects}
    \tiny
    \centering
    \setlength{\tabcolsep}{4pt}
    \resizebox{2\columnwidth}{!}{
    \begin{tabular}{@{}p{1.3cm}p{1.25cm}cccc@{\extracolsep{4pt}}cccc@{\extracolsep{2pt}}cccc@{}}
        \toprule
        \multirow{2}{*}{\shortstack{\textbf{Apache}\\ \textbf{projects}}} & \multirow{2}{*}{\shortstack{\textbf{Time}\\ \textbf{frame}$^\diamond$}}  &  \multicolumn{4}{c}{\textbf{Monthly data}} & \multicolumn{4}{c}{\textbf{Biweekly data}}\\
        \cmidrule{3-6} \cmidrule{7-10} \cmidrule{11-14}
         & & \textbf{ARIMAX} & \textbf{ARIMA+LM} & \textbf{SARIMAX} & \textbf{SARIMA+LM} & \textbf{ARIMAX} & \textbf{ARIMA+LM} & \textbf{SARIMAX} & \textbf{SARIMA+LM}\\
         \midrule
        \textbf{\textit{archiva}} & 05/11-02/12 & $\usym{1F5F8}$ & $\usym{1F5F8}$ & $\usym{1F5F8}$ & $\usym{1F5F8}$ & $\usym{1F5F8}$ & $\usym{1F5F8}$ & $\usym{1F5F8}$ & $\usym{1F5F8}$\\
        \textit{batik} & 10/00-08/02 & $\usym{1F5F8}$ & $\usym{1F5F8}$ & $\usym{1F5F8}$ & $\usym{1F5F8}$ &   &  &  & $\usym{1F5F8}$\\
        \textbf{\textit{bcel}} & 10/01-04/18 & $\usym{1F5F8}$ & $\usym{1F5F8}$ & $\usym{1F5F8}$ & $\usym{1F5F8}$ & $\usym{1F5F8}$ & $\usym{1F5F8}$ & $\usym{1F5F8}$ & $\usym{1F5F8}$\\
        \textit{beanutils} & 03/01-10/18 & $\usym{1F5F8}$ & $\usym{1F5F8}$ & $\usym{1F5F8}$ & $\usym{1F5F8}$ & $\usym{1F5F8}$ & $\usym{1F5F8}$ &  & $\usym{1F5F8}$\\
        \textit{cayenne} & 01/07-07/08 & $\usym{1F5F8}$ & $\usym{1F5F8}$ &  &  & $\usym{1F5F8}$ & $\usym{1F5F8}$ &  & \\
        \textit{cocoon} & 03/03-02/07 & $\usym{1F5F8}$ & $\usym{1F5F8}$ & $\usym{1F5F8}$ & $\usym{1F5F8}$ & $\usym{1F5F8}$ & $\usym{1F5F8}$ &  & $\usym{1F5F8}$\\
\end{tabular}
    }
\end{table*}

\begin{table}[h]
\caption{Summary statistics for the aggregated MAPE forecasting\\ results from the ARIMAX model with biweekly time series data}
\label{tab:biweeklyforecaststats}
\tiny
\begin{tabular}{crrrrr}
\hline
\textbf{Biweekly periods} & \textbf{Mean} & \textbf{Median} & \textbf{Max} & \textbf{Min} & \textbf{Variance} \\ \hline
1                                                                    & 1.44          & 0.12            & 17.82        & 0          & 13.06             \\
2                                                                    & 1.83          & 0.28            & 26.56        & 0          & 25.25             \\
3                                                                    & 2.14          & 0.31            & 28.83        & 0          & 29.38             \\
4                                                                    & 2.96          & 0.4             & 47.36        & 0          & 73.58             \\
5                                                                    & 3.51          & 0.47            & 58.31        & 0         & 110.16            \\
6                                                                    & 4.41          & 0.7             & 65.47        & 0          & 142.34            \\
7                                                                    & 5.09          & 0.87            & 70.46        & 0          & 171.62            \\
\hline
\end{tabular}
\end{table}

\begin{table}[]
\caption{Summary statistics for the aggregated MAPE forecasting\\ results from the SARIMAX model with monthly time series data}
\label{tab:monthlyforecaststats}
\scriptsize
\begin{tabular}{crrrrr}
\hline
\textbf{Monthly periods} & \textbf{Mean} & \textbf{Median} & \textbf{Max} & \textbf{Min} & \textbf{Variance} \\ \hline
1                                                                   & 0.3           & 0.08            & 1.52         & 0          & 0.18              \\
2                                                                   & 0.34          & 0.14            & 1.2          & 0          & 0.15              \\
3                                                                   & 0.5           & 0.25            & 2.1          & 0          & 0.32              \\
4                                                                   & 0.61          & 0.31            & 2.72         & 0          & 0.51              \\
5                                                                   & 0.73          & 0.39            & 3.32         & 0          & 0.76              \\
6                                                                   & 1.08          & 0.52            & 6.51         & 0          & 2.42              \\
7                                                                   & 1.38          & 0.57            & 10.37        & 0          & 5.36              \\
\hline
\end{tabular}
\end{table}

\end{document}